\documentclass[a4paper, 11pt]{article}

\usepackage{natbib}
\usepackage{amsmath}
\usepackage{amssymb}
\usepackage{amsthm}
\usepackage{hyperref}
\usepackage{graphicx}
\usepackage{color}
\usepackage{multirow}
\usepackage{booktabs}
\usepackage{pdfpages}
\usepackage[]{threeparttable}
\usepackage[left = 3cm, right = 2.5cm, bottom = 2cm, top = 3cm]{geometry}
\usepackage{tikz}
\usepackage{bm}

\newcommand*{\bb}{}

\def\E{{\rm E}}

\def\var{{\rm var}}

\def\btheta{\bb{\theta}}
\def\bbeta{\bb{\beta}}
\def\bgamma{\bb{\gamma}}
\def\bpsi{\bb{\psi}}
\def\blambda{\bb{\lambda}}
\def\bxi{\bb{\xi}}
\def\bX{\bb{X}}
\def\bx{\bb{x}}
\def\bW{\bb{W}}
\def\bD{\bb{D}}
\def\bH{\bb{H}}
\def\by{\bb{y}}
\def\bz{\bb{z}}
\def\bu{\bb{u}}
\def\bmu{\bb{\mu}}
\def\bs{\bb{s}}
\def\bi{\bb{i}}
\def\bzero{\bb{0}}
\def\bbias{\bb{b}}
\def\bA{\bb{A}}
\def\bB{\bb{B}}
\def\bK{\bb{K}}
\def\bM{\bb{M}}

\newtheoremstyle{example}
{3pt} 
{3pt} 
{} 
{0\parindent} 
{\bf}
{:} 
{.5em} 
{} 
\newtheoremstyle{theorem}
{3pt} 
{3pt} 
{\em} 
{0\parindent} 
{\bf}
{:} 
{.5em} 
{} 
\theoremstyle{example} \newtheorem{example}{Example}[section]
\theoremstyle{theorem} \newtheorem{theorem}{Theorem}[section]

\title{Mean and median bias reduction in generalized linear models}

\author{Ioannis Kosmidis \\
Department of Statistics \\
University of Warwick \\
Coventry, CV4 7AL, UK \smallskip \\
The Alan Turing Institute \\
96 Euston Road, London NW1 2DB, UK \\
\url{Ioannis.Kosmidis@warwick.ac.uk} \bigskip \\
Euloge Clovis Kenne Pagui \\
Department of Statistical Sciences \\
University of Padova \\
Via Cesare Battisti, 35121 Padova, Italy \\
\url{kenne@stat.unipd.it} \bigskip \\
Nicola Sartori \\
Department of Statistical Sciences \\
University of Padova \\
Via Cesare Battisti, 35121 Padova, Italy \\
\url{sartori@stat.unipd.it}
}

\begin{document}

\maketitle

\begin{abstract}
  This paper presents an integrated framework for estimation and
  inference from generalized linear models using adjusted score
  equations that result in mean and median bias reduction. The
  framework unifies theoretical and methodological aspects of past
  research on mean bias reduction and accommodates, in a natural way,
  new advances on median bias reduction. General expressions for the
  adjusted score functions are derived in terms of quantities that are
  readily available in standard software for fitting generalized
  linear models. The resulting estimating equations are solved using a
  unifying quasi-Fisher scoring algorithm that is shown to be
  equivalent to iteratively re-weighted least squares with
  appropriately adjusted working variates. Formal links between the
  iterations for mean and median bias reduction are established. Core
  model invariance properties are used to develop a novel mixed
  adjustment strategy when the estimation of a dispersion parameter is
  necessary. It is also shown how median bias reduction in multinomial
  logistic regression can be done using the equivalent Poisson
  log-linear model. The estimates coming out from mean and median bias
  reduction are found to overcome practical issues related to infinite
  estimates that can occur with positive probability in generalized
  linear models with multinomial or discrete responses, and can result
  in valid inferences even in the presence of a high-dimensional nuisance parameter. \\

\end{abstract}

\section{Introduction}
\label{model}

The flexibility of generalized linear models \citep{mccullagh+nelder:1989} in handling count, categorical, positive and real-valued responses under a common modelling framework has not only made them a typical choice in applications but also the focus of much methodological research on their estimation and use in inference.

Suppose that $y_1, \ldots, y_n$ are observations on independent random variables $Y_1, \ldots, Y_n$, each with probability density or mass function of the exponential family form
\[
f_{Y_i}(y; \theta_i, \phi) = \exp\left\{\frac{y \theta_i - b(\theta_i) - c_1(y)}{\phi/m_i} - \frac{1}{2}a\left(-\frac{m_i}{\phi}\right) + c_2(y) \right\}
\]
for some sufficiently smooth functions $b(\cdot)$, $c_1(\cdot)$, $a(\cdot)$ and $c_2(\cdot)$, and fixed observation weights $m_1, \ldots, m_n$. The expected value and the variance of $Y_i$ are then $\E(Y_i) = \mu_i = b'(\theta_i)$ and $\var(Y_i) = \phi b''(\theta_i)/m_i = \phi V(\mu_i)/m_i$, respectively, where  $b'(\theta_i)$ and $b''(\theta_i)$ are the first two derivatives of $b(\theta_i)$. Compared to the normal distribution, exponential family models are generally  heteroscedastic because the response variance depends on the mean through the variance function $V(\mu_i)$, and the dispersion parameter $\phi$ allows shrinking or inflating that contribution of the mean. A generalized linear model (GLM) links the mean $\mu_i$ to a linear predictor $\eta_i$ through a monotone, sufficiently smooth link function $g(\mu_i) = \eta_i$ with $\eta_i = \sum_{t=1}^p \beta_t x_{it}$ where $x_{it}$ is the $(i,t)th$ component of a model matrix $\bX$, and $\bbeta = (\beta_1, \ldots, \beta_p)^\top$. An intercept parameter is typically included in the linear predictor, in which case $x_{i1} = 1$ for all $i \in \{1, \ldots, n\}$.

Estimation of the parameters of GLMs is commonly done using maximum likelihood (ML) because of the limiting guarantees that the ML estimator provides assuming that the model assumptions are adequate. Specifically, the ML estimator $(\hat\bbeta^\top, \hat\phi)^\top$ is consistent, asymptotically unbiased and asymptotically efficient with a limiting normal distribution centred at the target parameter value and a variance-covariance matrix, given by the inverse of the Fisher information matrix, which is also the Cram\'{e}r-Rao lower bound for the variance of unbiased estimators. These properties are used as re-assurance that inferential procedures based on Wald, score or likelihood ratio statistics will perform well in large samples.  Another reason that ML is the default estimation method for GLMs is that maximizing the likelihood can be conveniently performed by iteratively reweighted least squares \citep[IWLS;][]{green:1984}, requiring only standard algorithms for least squares and the evaluation of working weights and variates at each iteration.

Nevertheless, the properties of the ML estimator and of the associated inferential procedures that depend on its asymptotic normality may deteriorate for small or moderate sample sizes or, more generally, when the number of parameters is large relative to the number of observations.

\begin{example}
  \label{clotting}
  To illustrate the differences between finite-sample and limiting behaviour of the ML estimator and associate inferential procedures, consider the data in \citet[\S~8.4.2]{mccullagh+nelder:1989} of mean blood clotting times in seconds for nine percentage concentrations of normal plasma and two lots of clotting agent. The plasma concentrations are 5, 10, 15,  20, 30, 40, 60, 80, 100, with corresponding clotting times 118, 58, 42, 35, 27, 25, 21, 19, 18 for the first lot, and 69, 35, 26, 21, 18, 16, 13, 12, 12 for the second lot, respectively.  We fit a Gamma GLM with $\log \mu_i = \sum_{t=1}^4 \beta_t x_{it}$, where $\mu_i$ is the expectation of the $i$th clotting time, $x_{i1}=1$, $x_{i2}$ is $1$ for the second lot and $0$ otherwise, $x_{i3}$ is the corresponding (log) plasma concentration, and $x_{i4}=x_{i2}x_{i3}$ is an interaction term. The ML estimates are $\hat\bbeta=(5.503, -0.584, -0.602, 0.034)$ and $\hat\phi=0.017$. Table~\ref{tab:estimators2.1} shows the estimated bias, root mean squared error, percentage of underestimation and mean absolute error of the ML estimator from $10\, 000$ simulated samples at the ML estimates, with covariates values fixed as in the original sample. The table also includes the same summaries of the moment-based estimator of $\phi$ (see, for example, \citealt{mccullagh+nelder:1989}, \S~8.3, and the \texttt{summary.glm} function in R). The ML estimator of the regression parameters illustrates good bias properties, with distributions that have a mode around the parameter value used for simulation. On the other hand, the ML estimator of the dispersion parameter is subject to severe bias, which inflates the mean squared error by $54.13\%$ from its absolute minimum, and has a severely right skewed distribution. Note here that the latter observation holds for any monotone transformation of the dispersion parameter. The moment-based estimator on the other hand has a much smaller bias, probability of underestimation closer to $0.5$, and its use delivers a marked improvement to the coverage of standard  confidence intervals for all model parameters.
\end{example}

\begin{table}[t!]
  \caption{Clotting data. Estimated bias (B), root mean squared error (RMSE), percentage of underestimation (PU), mean absolute error (MAE) of maximum likelihood estimator, and coverage of nominally $95\%$ Wald-type confidence intervals (C), based on $10\, 000$ samples under the ML fit. The summary $\text{B}^2/\text{SD}^2$ is the relative increase in mean squared error from its absolute minimum due to bias.  The results include the same summaries of the moment-based estimator of $\phi$ (row marked with $\star$). All reported figures are    $\times 100$ of their actual value and $<0.01$ is used for a value that is less than $0.01$ in absolute value.}
  \begin{center}
    \begin{tabular}{crrrrrr}
      \toprule
      \multicolumn{1}{c}{Parameter} & \multicolumn{1}{c}{B} & \multicolumn{1}{c}{RMSE} & \multicolumn{1}{c}{$\text{B}^2/\text{SD}^2$} & \multicolumn{1}{c}{PU} & \multicolumn{1}{c}{MAE} & \multicolumn{1}{c}{C} \\ \midrule
      $\beta_1$ & -0.33 & 16.15 & 0.04 & 50.42 & 12.87 & 89.26 \\
                                    & & & & & & $^\star$93.05 \\
      $\beta_2$ & 0.36 & 23.09 & 0.02 & 49.61 & 18.46 & 88.87 \\
                                    & & & & & & $^\star$92.66 \\
      $\beta_3$ & 0.06 & 4.69 & 0.01 & 49.73 & 3.74 & 89.62 \\
                                    & & & & & & $^\star$93.04 \\
      $\beta_4$ & -0.11 & 6.71 & 0.03 & 50.51 & 5.36 & 88.78  \\
                                    & & & & & & $^\star$92.47\\
      $\phi$ & -0.38 & 0.65 & 54.13 & 78.77 & 0.55 & \\
             & $^\star$$<$0.01 & $^\star$0.67&  $^\star$$<$0.01 & $^\star$55.61 & $^\star$0.53 & \\ \bottomrule
    \end{tabular}
  \end{center}
  \label{tab:estimators2.1}
\end{table}

Improvements of first-order inference based on ML can be achieved in several ways. For instance, bootstrap methods guarantee both correction of bias and higher-order accurate inference. Alternatively, analytical methods derived from higher-order asymptotic expansions based on the likelihood (see for instance \citealt{brazzale+davison+reid:2007}) have been found to result in accurate inference on model parameters. Nevertheless, bootstrap methods typically require intensive computation, and analytical methods, typically, require tedious, model-specific algebraic effort for their implementation. Furthermore, both bootstrap and analytical methods rely on the existence of the ML estimate, which is not always guaranteed. Such an example is GLMs with multinomial or discrete responses \citep{heinze+schemper:2002, kosmidis:2014b}.

This paper presents a unified approach for mean and median bias reduction (BR) in GLMs using adjusted score functions (\citealt{firth:1993, kosmidis+firth:2009}, and \citealt{kenne+salvan+sartori:2017}, respectively). Specifically, \citet{firth:1993} and \citet{kosmidis+firth:2009} achieve higher-order BR of the ML estimator through the additive adjustment of the score equation. \citet{kenne+salvan+sartori:2017} use a similar approach in order to obtain component-wise higher-order median BR of the ML estimator, i.e.~each component of the estimator has, to third-order, the same probability of underestimating and overestimating the corresponding parameter component. We illustrate how those methods can be implemented  without sacrificing the computational simplicity and the first-order inferential properties of the ML framework, and illustrate that they provide simple and practical solutions to the issue of boundary estimates in models with categorical responses.

Explicit, general formulae are derived for the adjusted score equations that produce higher-order mean and median unbiased estimators for GLMs. It is shown that, like ML, both mean and median BR can be conveniently performed by IWLS after the appropriate adjustment of the working variates for ML. Extensive empirical evidence illustrate that such an adjustment of IWLS leads to a stable estimation procedure even in case in which standard IWLS for ML estimation diverges.

Each method possesses invariance properties that can be more useful or less desirable depending on the GLM under consideration; the estimators resulting from mean BR (mean BR estimators, in short) are exactly invariant under linear transformations of the parameters in terms of the mean bias of the transformed estimators, which is useful, for example, when estimation and inference on arbitrary contrasts of the regression parameters is of interest. These invariance properties do not extend, though, to more general nonlinear transformations. On the other hand, median BR delivers estimators that are exactly invariant in terms of their improved median bias properties under general component-wise transformations of the parameters, which is useful, for example, when a dispersion parameter needs to be estimated from data. However, estimators from median BR are not invariant in terms of the median bias properties under more general transformations, like for example, parameter contrasts. In order to combine the desirable invariance properties of each method when modelling with GLMs, we exploit the Fisher orthogonality \citep{cox+reid:1987} of the mean and dispersion parameters to formally derive a novel mixed adjustment approach that delivers estimators of the regression parameters with improved mean bias, and estimators for any unknown dispersion parameter with improved median bias.

Examples and simulation studies for various response distributions are used to demonstrate that both methods for BR are effective in achieving their respective goals and improve upon maximum likelihood, even in extreme settings characterized by high-dimensional nuisance parameters. Particular focus is given on special cases, like estimation of odds-ratios from logistic regression models and estimation of log-odds ratios from multinomial baseline category models.

All methods and algorithms discussed in this paper are implemented in the \texttt{brglm2} R package \citep{brglm2}, which has been used for all numerical computations and simulation experiments (see Supplementary Material).

The remaining of the paper is structured as follows. Section~\ref{sec:br_iwsm} gives a brief introduction to estimation using IWLS, and shows how IWLS can be readily adjusted to perform mean or median BR.  In particular, Subsections \ref{sec:iwls} and \ref{explicit} review known results for ML estimation and explicit, mean bias correction in generalized linear models. These subsections are useful to setup the notation and allow the introduction of mean and median bias-reducing adjusted score functions in Subsections \ref{meanadjustedscores} and \ref{medianadjustedscores} respectively. Inferential procedures based on the bias-reduced estimators are discussed in Section~\ref{sec:br_inference}. Section~\ref{sec:br_mixed} motivates the need for and introduces the mixed adjustment strategy for GLMs with a dispersion parameter. All methods are then assessed and compared through case studies and simulation experiments in Section~\ref{sec:br_illustrations} and Section~\ref{sec:multinomial}. Section~\ref{sec:multinomial} also discusses how multinomial logistic regression models can be easily estimated with all methods using the equivalent Poisson log-linear model. Section~\ref{sec:discussion} concludes the paper with a short discussion and possible extensions.


\section{Bias reduction and iteratively reweighted least squares}
\label{sec:br_iwsm}
\subsection{Iteratively reweighted least squares}
\label{sec:iwls}

The log-likelihood function for a GLM is $\sum_{i = 1}^n \log f_{Y_i}(y_i; g^{-1}(\eta_i), \phi)$, where $g^{-1}(\cdot)$ is the inverse of the link function. Suppressing the dependence of the various quantities on the model parameters and the data, the derivatives of the log-likelihood function with respect to the components of $\bbeta$ and $\phi$ are
\begin{equation}
  \label{scores} \bs_{\bbeta} = \frac{1}{\phi}\bX^T\bW\bD^{-1}(\by - \bmu) \quad \text{and} \quad s_\phi = \frac{1}{2\phi^2}\sum_{i = 1}^n (q_i - \rho_i) \, ,
\end{equation}
respectively, with $\by = (y_1, \ldots, y_n)^\top$, $\bmu = (\mu_1, \ldots, \mu_n)^\top$, $\bW = {\rm diag}\left\{w_1, \ldots, w_n\right\}$ and $\bD = {\rm diag}\left\{d_1, \ldots, d_n\right\}$, where $w_i = m_i d_i^2/v_i$ is the $i$th working weight, with $d_i = d\mu_i/d\eta_i$ and $v_i = V(\mu_i)$. Furthermore, $q_i = -2 m_i \{y_i\theta_i - b(\theta_i) - c_1(y_i)\}$ and $\rho_i = m_i a'_i$ are the $i$th deviance residual and its expectation, respectively, with $a'_i = a'(-m_i/\phi)$, where $a'(u) = d a(u)/d u$.

The ML estimators $\hat\bbeta$ of $\bbeta$ and $\hat\phi$ of $\phi$, can be found by solution of the score equations $\bs_{\bbeta} = 0_p$ and $s_\phi = 0$, where $0_p$ is a $p$-dimensional vector of zeros. \citet{wedderburn:1976} derives necessary and sufficient conditions for the existence and uniqueness of the ML estimator of $\hat\bbeta$. Given that the dispersion parameter $\phi$ appears in the expression for $\bs_{\bbeta}$ in~(\ref{scores}) only multiplicatively, the ML estimate of $\bbeta$ can be computed without knowledge of the value of $\phi$. This fact is exploited in popular software like the \texttt{glm.fit} function in R \citep{rproject}. The $j$th iteration of IWLS updates the current iterate $\bbeta^{(j)}$ for $\bbeta$ by solving the weighted least squares problem
\begin{equation}
  \label{iwls}
  \left(\bX^\top \bW^{(j)} \bX\right)^{-1} \bX^\top \bW^{(j)}\bz^{(j)}\,,
\end{equation}
where the superscript $(j)$ indicates evaluation at $\bbeta^{(j)}$, and $\bz = (z_1, \ldots, z_n)^\top$ is the vector of ``working'' variates with $z_i = \eta_i + (y_i - \mu_i)/d_i$ \citep{green:1984}. Table \ref{adjusted_variates} reports the working variates for well-used combinations of exponential family models and link functions. The updated $\bbeta$ from the weighted least squares problem in (\ref{iwls}) is equal to the updated $\bbeta$ from the Fisher scoring step
\[
  \bbeta^{(j)} + \left\{\bi_{\bbeta\bbeta}^{(j)}\right\}^{-1} \bs_{\bbeta}^{(j)}\,,
\]
where $\bi_{\bbeta\bbeta}$ is the $(\bbeta,\bbeta)$ block of the expected information matrix about $\bbeta$ and $\phi$
\begin{equation}
  \label{information}
\bi =
\left[
\begin{array}{cc}
\bi_{\bbeta\bbeta} & \bzero_p \\
\bzero_p^\top & i_{\phi\phi}
\end{array}
\right]
=
\left[
\begin{array}{cc}
\frac{1}{\phi} \bX^\top \bW \bX & \bzero_p \\
\bzero_p^\top & \frac{1}{2\phi^4}\sum_{i = 1}^n m_i^2 a''_i
\end{array}
\right]\,,
\end{equation}
with  $a''_i = a''(-m_i/\phi)$, where $a''(u) = d^2 a(u)/d u^2$.

\subsection{Explicit mean bias reduction}
\label{explicit}
\citet{efron:1975} has shown that under the usual regularity conditions, the asymptotic mean bias of the ML estimator $\hat\bgamma$ for a general parametric model $\mathcal{M}_{\bgamma}$ can be reduced by the explicit correction of $\hat\bgamma$ as $\tilde\bgamma = \hat\bgamma - \bbias_{\bgamma}(\hat\bgamma)$, where $\bbias_{\bgamma} \equiv \bbias_{\bgamma}(\bgamma)$ is the first term in the expansion of the mean bias of $\hat\bgamma$. \citet{kosmidis:2014a} provides a review of explicit and implicit methods for mean BR. The general form of $\bbias_{\bgamma}$ is given in \citet{cox+snell:1968} in index notation and in \citet[Section~2]{kosmidis+firth:2010} in matrix notation. For GLMs, $\bbias_{\bbeta} = -\bi_{\bbeta\bbeta}^{-1} \bA_{\bbeta}^*$ and $b_\phi = -i_{\phi\phi}^{-1} A_\phi^*$ with
\begin{equation}
  \label{mean_adjustments}
  \bA_{\bbeta}^*  = \bX^\top \bW \bxi \quad \text{and} \quad A_\phi^* = \frac{(p - 2)}{2\phi} + \frac{\sum_{i = 1}^n m_i^3 a'''_i}{2\phi^2\sum_{i = 1}^n m_i^2 a''_i}\,,
\end{equation}
where $\bxi = (\xi_1, \ldots, \xi_n)^T$ with $\xi_i = h_id_i'/(2d_iw_i)$ and $d_i' = d^2\mu_i/d\eta_i^2$, $h_i$ is the ``hat'' value for the $i$th observation, obtained as the $i$th diagonal element of the matrix $\bH = \bX (\bX^\top \bW \bX)^{-1} \bX^\top \bW$, and $a'''_i = a'''(-m_i/\phi)$, with $a'''(u) = d^3 a(u)/d u^3$. The derivation of $b_\phi$ above is done using \citet[expressions~(4.8) in Remark 3]{kosmidis+firth:2010} to write $b_\phi$ in terms of the first term in the expansion of the bias of $1/\hat{\phi}$, which is given in \cite{cordeiro+mccullagh:1991}.

Note here that neither $i_{\phi\phi}$ nor $A_\phi^*$ depend on $\bbeta$
and hence the bias-reduced estimator for $\phi$ can be computed by
knowledge of $\hat\phi$ only as
\[
\hat\phi \left\{ 1 + \hat\phi \frac{\sum m_i^3
    \hat{a}_i'''}{\left(\sum m_i^2 \hat{a}_i''\right)^2} + \hat\phi^2 \frac{p - 2}{\sum m_i^2
    \hat{a}_i''}
\right\} \, ,
\]
where $\hat{a}_i''' = a'''(-m_i/\hat\phi)$. Some algebra gives that
the bias-reduced estimator for $\bbeta$ is
\begin{equation}
  \label{adjusted_iwls}
  \left(\bX^\top \hat{\bW} \bX\right)^{-1} \bX^\top \hat{\bW}
  \left(\hat{\bz} + \hat{\phi}\hat{\bxi}
  \right) \, ,
\end{equation}
where $\hat{\bB}$ denotes evaluation of $\bB$ at the ML
estimator. Equivalently, and as also noted in
\citet{cordeiro+mccullagh:1991}, the explicit correction
$\hat\bbeta - \bbias_{\bbeta}(\hat\bbeta, \hat\phi)$ can be performed by IWLS as
in~(\ref{iwls}) up to convergence, and then making one extra step,
where the working variate $\bz$ is replaced by its adjusted version
$\bz + \phi\bxi$.  Table \ref{adjusted_variates} gives the quantity
$\phi\bxi$ for some well-used GLMs.

\subsection{Mean bias-reducing adjusted score functions}
\label{meanadjustedscores}
\citet{firth:1993} shows that the solution of the adjusted score equations
\begin{equation}
  \label{AS_mean}
  \bs_{\bbeta} + \bA_{\bbeta}^* = \bzero_p \quad \text{and} \quad s_\phi + A_\phi^* = 0
\end{equation}
with $\bA_{\bbeta}^*$ and $A_\phi^*$ as in (\ref{mean_adjustments}) result in
estimators $\bbeta^*$ and $\phi^*$ with mean bias of smaller
asymptotic order than the ML estimator.

\begin{figure}[t]
  \caption{Illustration of the quasi-Fisher scoring iterations for a
    model with a scalar parameter $\beta$, starting at the maximum
    likelihood estimate $\hat\beta$. One step gives the explicit mean
    reduced-bias estimator $\hat\beta - b_{\beta}(\hat\beta)$ of
    Section~\ref{explicit}, and iterating until convergence results in
    the solution $\beta^*$ of the mean bias-reducing adjusted score
    equation.}
  \begin{center}
    \ifx\du\undefined
  \newlength{\du}
\fi
\setlength{\du}{15\unitlength}
\begin{tikzpicture}
\pgftransformxscale{1.000000}
\pgftransformyscale{-1.000000}
\definecolor{dialinecolor}{rgb}{0.000000, 0.000000, 0.000000}
\pgfsetstrokecolor{dialinecolor}
\definecolor{dialinecolor}{rgb}{1.000000, 1.000000, 1.000000}
\pgfsetfillcolor{dialinecolor}
\pgfsetlinewidth{0.100000\du}
\pgfsetdash{}{0pt}
\pgfsetdash{}{0pt}
\pgfsetmiterjoin
\pgfsetbuttcap
{
\definecolor{dialinecolor}{rgb}{0.749020, 0.749020, 0.749020}
\pgfsetfillcolor{dialinecolor}
\definecolor{dialinecolor}{rgb}{0.749020, 0.749020, 0.749020}
\pgfsetstrokecolor{dialinecolor}
\pgfpathmoveto{\pgfpoint{35.259932\du}{21.830667\du}}
\pgfpathcurveto{\pgfpoint{41.459932\du}{21.780667\du}}{\pgfpoint{45.832122\du}{37.439385\du}}{\pgfpoint{55.832122\du}{38.439385\du}}
\pgfusepath{stroke}
}
\pgfsetlinewidth{0.100000\du}
\pgfsetdash{}{0pt}
\pgfsetdash{}{0pt}
\pgfsetmiterjoin
\pgfsetbuttcap
{
\definecolor{dialinecolor}{rgb}{0.749020, 0.749020, 0.749020}
\pgfsetfillcolor{dialinecolor}
\definecolor{dialinecolor}{rgb}{0.749020, 0.749020, 0.749020}
\pgfsetstrokecolor{dialinecolor}
\pgfpathmoveto{\pgfpoint{33.000000\du}{40.000000\du}}
\pgfpathcurveto{\pgfpoint{33.700000\du}{35.750000\du}}{\pgfpoint{56.497936\du}{27.056828\du}}{\pgfpoint{57.547936\du}{20.256828\du}}
\pgfusepath{stroke}
}
\pgfsetlinewidth{0.100000\du}
\pgfsetdash{{1.000000\du}{1.000000\du}}{0\du}
\pgfsetdash{{1.000000\du}{1.000000\du}}{0\du}
\pgfsetbuttcap
{
\definecolor{dialinecolor}{rgb}{0.749020, 0.749020, 0.749020}
\pgfsetfillcolor{dialinecolor}
\definecolor{dialinecolor}{rgb}{0.749020, 0.749020, 0.749020}
\pgfsetstrokecolor{dialinecolor}
\draw (34.127658\du,33.807109\du)--(55.653941\du,33.802442\du);
}
\pgfsetlinewidth{0.100000\du}
\pgfsetdash{}{0pt}
\pgfsetdash{}{0pt}
\pgfsetbuttcap
{
\definecolor{dialinecolor}{rgb}{0.749020, 0.749020, 0.749020}
\pgfsetfillcolor{dialinecolor}
\definecolor{dialinecolor}{rgb}{0.749020, 0.749020, 0.749020}
\pgfsetstrokecolor{dialinecolor}
\draw (45.173104\du,34.127462\du)--(45.173104\du,33.528588\du);
}
\pgfsetlinewidth{0.100000\du}
\pgfsetdash{}{0pt}
\pgfsetdash{}{0pt}
\pgfsetbuttcap
{
\definecolor{dialinecolor}{rgb}{0.749020, 0.749020, 0.749020}
\pgfsetfillcolor{dialinecolor}
\definecolor{dialinecolor}{rgb}{0.749020, 0.749020, 0.749020}
\pgfsetstrokecolor{dialinecolor}
\draw (47.568192\du,34.117073\du)--(47.568192\du,33.518199\du);
}
\definecolor{dialinecolor}{rgb}{0.000000, 0.000000, 0.000000}
\pgfsetstrokecolor{dialinecolor}
\node[anchor=west] at (40.388740\du,24.456558\du){$s_{\beta}(\beta)/i_{\beta\beta}(\beta)$};
\definecolor{dialinecolor}{rgb}{0.000000, 0.000000, 0.000000}
\pgfsetstrokecolor{dialinecolor}
\node[anchor=east] at (53.942945\du,24.466602\du){$b_\beta(\beta)$};
\pgfsetlinewidth{0.050000\du}
\pgfsetdash{}{0pt}
\pgfsetdash{}{0pt}
\pgfsetbuttcap
{
\definecolor{dialinecolor}{rgb}{0.000000, 0.000000, 0.000000}
\pgfsetfillcolor{dialinecolor}
\pgfsetarrowsstart{stealth}
\definecolor{dialinecolor}{rgb}{0.000000, 0.000000, 0.000000}
\pgfsetstrokecolor{dialinecolor}
\draw (47.575115\du,33.734748\du)--(47.592989\du,29.605855\du);
}
\pgfsetlinewidth{0.050000\du}
\pgfsetdash{}{0pt}
\pgfsetdash{}{0pt}
\pgfsetbuttcap
{
\definecolor{dialinecolor}{rgb}{0.000000, 0.000000, 0.000000}
\pgfsetfillcolor{dialinecolor}
\pgfsetarrowsstart{stealth}
\definecolor{dialinecolor}{rgb}{0.000000, 0.000000, 0.000000}
\pgfsetstrokecolor{dialinecolor}
\draw (43.017246\du,28.372549\du)--(42.999372\du,32.304828\du);
}
\pgfsetlinewidth{0.050000\du}
\pgfsetdash{}{0pt}
\pgfsetdash{}{0pt}
\pgfsetbuttcap
{
\definecolor{dialinecolor}{rgb}{0.000000, 0.000000, 0.000000}
\pgfsetfillcolor{dialinecolor}
\pgfsetarrowsstart{stealth}
\definecolor{dialinecolor}{rgb}{0.000000, 0.000000, 0.000000}
\pgfsetstrokecolor{dialinecolor}
\draw (46.806533\du,32.930418\du)--(46.806533\du,30.142075\du);
}
\pgfsetlinewidth{0.050000\du}
\pgfsetdash{}{0pt}
\pgfsetdash{}{0pt}
\pgfsetbuttcap
{
\definecolor{dialinecolor}{rgb}{0.000000, 0.000000, 0.000000}
\pgfsetfillcolor{dialinecolor}
\pgfsetarrowsstart{stealth}
\definecolor{dialinecolor}{rgb}{0.000000, 0.000000, 0.000000}
\pgfsetstrokecolor{dialinecolor}
\draw (44.018190\du,29.641603\du)--(44.000316\du,31.697113\du);
}
\pgfsetlinewidth{0.050000\du}
\pgfsetdash{}{0pt}
\pgfsetdash{}{0pt}
\pgfsetbuttcap
{
\definecolor{dialinecolor}{rgb}{0.000000, 0.000000, 0.000000}
\pgfsetfillcolor{dialinecolor}
\pgfsetarrowsstart{stealth}
\definecolor{dialinecolor}{rgb}{0.000000, 0.000000, 0.000000}
\pgfsetstrokecolor{dialinecolor}
\draw (46.070357\du,32.110986\du)--(46.091573\du,30.571051\du);
}
\pgfsetlinewidth{0.050000\du}
\pgfsetdash{{\pgflinewidth}{0.200000\du}}{0cm}
\pgfsetdash{{\pgflinewidth}{0.200000\du}}{0cm}
\pgfsetbuttcap
{
\definecolor{dialinecolor}{rgb}{0.000000, 0.000000, 0.000000}
\pgfsetfillcolor{dialinecolor}
\definecolor{dialinecolor}{rgb}{0.000000, 0.000000, 0.000000}
\pgfsetstrokecolor{dialinecolor}
\draw (47.503619\du,33.786912\du)--(42.960267\du,33.786107\du);
}
\pgfsetlinewidth{0.050000\du}
\pgfsetdash{{\pgflinewidth}{0.200000\du}}{0cm}
\pgfsetdash{{\pgflinewidth}{0.200000\du}}{0cm}
\pgfsetbuttcap
{
\definecolor{dialinecolor}{rgb}{0.000000, 0.000000, 0.000000}
\pgfsetfillcolor{dialinecolor}
\definecolor{dialinecolor}{rgb}{0.000000, 0.000000, 0.000000}
\pgfsetstrokecolor{dialinecolor}
\draw (46.766504\du,28.352921\du)--(42.978957\du,28.365385\du);
}
\pgfsetlinewidth{0.050000\du}
\pgfsetdash{{\pgflinewidth}{0.200000\du}}{0cm}
\pgfsetdash{{\pgflinewidth}{0.200000\du}}{0cm}
\pgfsetbuttcap
{
\definecolor{dialinecolor}{rgb}{0.000000, 0.000000, 0.000000}
\pgfsetfillcolor{dialinecolor}
\definecolor{dialinecolor}{rgb}{0.000000, 0.000000, 0.000000}
\pgfsetstrokecolor{dialinecolor}
\draw (46.771867\du,32.951952\du)--(44.083156\du,32.962655\du);
}
\pgfsetlinewidth{0.050000\du}
\pgfsetdash{{\pgflinewidth}{0.200000\du}}{0cm}
\pgfsetdash{{\pgflinewidth}{0.200000\du}}{0cm}
\pgfsetbuttcap
{
\definecolor{dialinecolor}{rgb}{0.000000, 0.000000, 0.000000}
\pgfsetfillcolor{dialinecolor}
\definecolor{dialinecolor}{rgb}{0.000000, 0.000000, 0.000000}
\pgfsetstrokecolor{dialinecolor}
\draw (46.037951\du,29.628079\du)--(44.002149\du,29.633324\du);
}
\pgfsetlinewidth{0.050000\du}
\pgfsetdash{{\pgflinewidth}{0.200000\du}}{0cm}
\pgfsetdash{{\pgflinewidth}{0.200000\du}}{0cm}
\pgfsetbuttcap
{
\definecolor{dialinecolor}{rgb}{0.000000, 0.000000, 0.000000}
\pgfsetfillcolor{dialinecolor}
\definecolor{dialinecolor}{rgb}{0.000000, 0.000000, 0.000000}
\pgfsetstrokecolor{dialinecolor}
\draw (46.021528\du,32.134755\du)--(44.502620\du,32.140000\du);
}
\pgfsetlinewidth{0.050000\du}
\pgfsetdash{}{0pt}
\pgfsetdash{}{0pt}
\pgfsetbuttcap
{
\definecolor{dialinecolor}{rgb}{0.000000, 0.000000, 0.000000}
\pgfsetfillcolor{dialinecolor}
\pgfsetarrowsstart{stealth}
\definecolor{dialinecolor}{rgb}{0.000000, 0.000000, 0.000000}
\pgfsetstrokecolor{dialinecolor}
\draw (44.502620\du,30.348788\du)--(44.496797\du,31.443577\du);
}
\definecolor{dialinecolor}{rgb}{0.000000, 0.000000, 0.000000}
\pgfsetstrokecolor{dialinecolor}
\node[anchor=west] at (46.733199\du,34.934939\du){$\hat\beta$};
\definecolor{dialinecolor}{rgb}{0.000000, 0.000000, 0.000000}
\pgfsetstrokecolor{dialinecolor}
\node[anchor=west] at (44.431007\du,34.936557\du){$\beta^*$};
\definecolor{dialinecolor}{rgb}{0.000000, 0.000000, 0.000000}
\pgfsetstrokecolor{dialinecolor}
\node[anchor=west] at (40.714489\du,24.204122\du){};
\definecolor{dialinecolor}{rgb}{0.000000, 0.000000, 0.000000}
\pgfsetstrokecolor{dialinecolor}
\node[anchor=west] at (53.540375\du,24.254029\du){};
\pgfsetlinewidth{0.050000\du}
\pgfsetdash{{\pgflinewidth}{0.200000\du}}{0cm}
\pgfsetdash{{\pgflinewidth}{0.200000\du}}{0cm}
\pgfsetbuttcap
{
\definecolor{dialinecolor}{rgb}{0.000000, 0.000000, 0.000000}
\pgfsetfillcolor{dialinecolor}
\definecolor{dialinecolor}{rgb}{0.000000, 0.000000, 0.000000}
\pgfsetstrokecolor{dialinecolor}
\draw (45.630247\du,30.317628\du)--(44.514955\du,30.320489\du);
}
\pgfsetlinewidth{0.100000\du}
\pgfsetdash{}{0pt}
\pgfsetdash{}{0pt}
\pgfsetbuttcap
{
\definecolor{dialinecolor}{rgb}{0.749020, 0.749020, 0.749020}
\pgfsetfillcolor{dialinecolor}
\definecolor{dialinecolor}{rgb}{0.749020, 0.749020, 0.749020}
\pgfsetstrokecolor{dialinecolor}
\draw (42.992610\du,34.109514\du)--(42.992610\du,33.510640\du);
}
\definecolor{dialinecolor}{rgb}{0.000000, 0.000000, 0.000000}
\pgfsetstrokecolor{dialinecolor}
\node[anchor=west] at (40.250513\du,34.918609\du){$\hat\beta - b_{\beta}(\hat\beta)$};
\definecolor{dialinecolor}{rgb}{0.000000, 0.000000, 0.000000}
\pgfsetstrokecolor{dialinecolor}
\node[anchor=west] at (42.533836\du,34.683851\du){};
\end{tikzpicture}

  \end{center}
  \label{quasi_figure}
\end{figure}
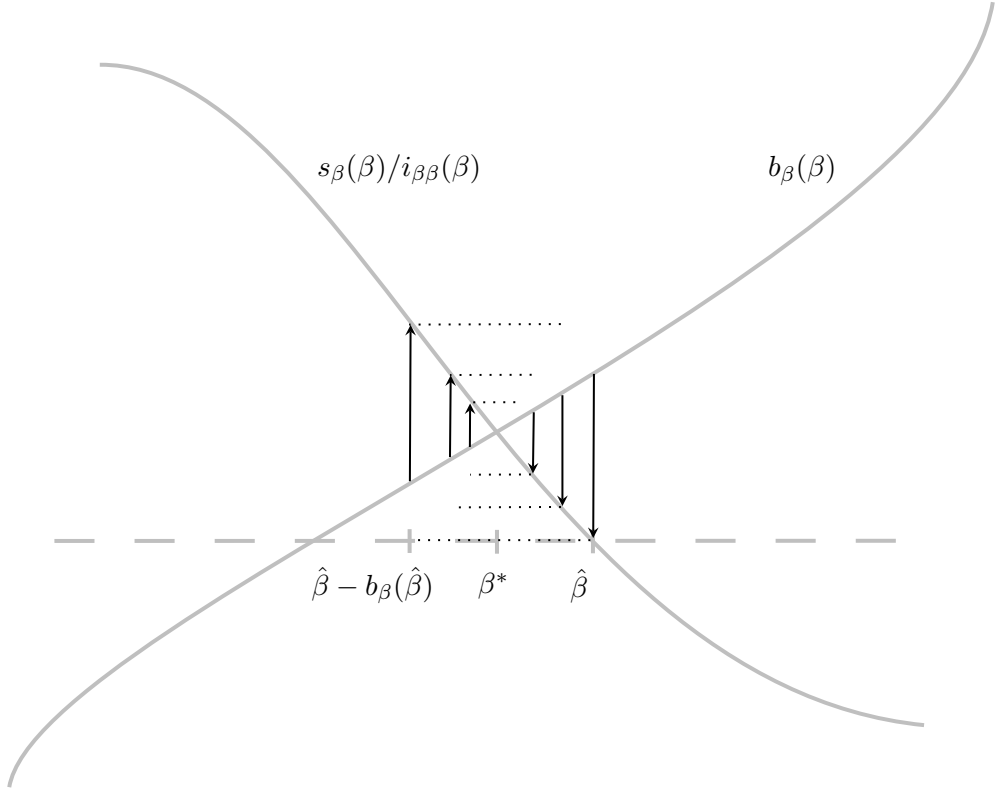

A natural way to solve the adjusted score equations is through
quasi-Fisher scoring \citep[see,][for the corresponding quasi
Newton-Raphson iteration]{kosmidis+firth:2010}, where at the $j$th
step the values for $\bbeta$ and $\phi$ are updated as
\begin{align}
  \notag
  \bbeta^{(j+1)} & \leftarrow \bbeta^{(j)} + \left\{\bi_{\bbeta\bbeta}^{(j)}\right\}^{-1}
                  \bs_{\bbeta}^{(j)} - \bbias_{\bbeta}^{(j)}\,, \\
  \label{mean_qfs}
  \phi^{(j + 1)} & \leftarrow \phi^{(j)} +
                   \left\{i_{\phi\phi}^{(j)}\right\}^{-1}s_{\phi}^{(j)}
                   - b_\phi^{(j)} \, .
\end{align}
The term ``quasi'' here reflects the fact that the expectation of the
negative second derivatives of the scores, instead of the adjusted
scores, is used for the calculation of the step size. Setting
$\phi^{(j + 1)} - \phi^{(j)} = 0$ in the above iteration shows that it
has the required stationary point. Furthermore, if the starting values
$\bbeta^{(0)}$ and $\phi^{(0)}$ for iteration (\ref{mean_qfs}) are the
ML estimates, then $\bbeta^{(1)}$ and $\phi^{(1)}$ are the estimates
from explicit BR, because $\bs_{\bbeta}^{(0)} = \bzero_p$ and
$s_{\phi}^{(0)} = 0$. Figure~\ref{quasi_figure} illustrates the
quasi-Fisher scoring iterations for an one-parameter problem, starting
from the ML estimate.

\begin{table}[t!]
  \caption{Working variates for ML, and additional quantities needed in mean and median BR, for the most popular combinations of distributions and link functions.}
  \begin{center}
    \resizebox{\textwidth}{!}{
      \begin{tabular}{cccccc}
        \toprule
        Distribution & $\eta$ & ML & mean BR & median BR \\ \cmidrule{3-5}
                     & $ $ & $\eta+(y-\mu)/d$ & $\phi\xi$ & $d v' / (6 v) - d'/(2 d)$  \\ \midrule
        Normal & $\mu$ & $y$ & 0 & 0  \bigskip \\
        Binomial & $\displaystyle \log\frac{\mu}{1-\mu}$ & $\displaystyle \eta+\frac{y-\mu}{\mu(1-\mu)}$ &  $\displaystyle \frac{h\{e^\eta-e^{-\eta}\}}{2m}$ & $\displaystyle \frac{2(1-e^\eta)}{3(1+e^\eta)}$ \medskip \\
                     & $\displaystyle \Phi^{-1}(\mu)$ & $\displaystyle \eta + \frac{y-\mu}{\phi(\eta)}$ & $\displaystyle -\frac{h\eta\{\Phi(\eta)(1-\Phi(\eta))\}}{2m\phi(\eta)^2}$ & $\displaystyle \frac{\phi(\eta)(1-2\Phi(\eta))}{6\Phi(\eta)(1-\Phi(\eta))}+\frac{\eta}{2}$ \medskip \\
                     & $\displaystyle \log\{-\log(1 - \mu)\}$ & $\displaystyle \eta + \frac{y-\mu}{e^{\eta-e^\eta}}$ & $\displaystyle \frac{h\mu\{1-e^\eta\}}{2me^{2\eta-e^\eta}}$ & $\displaystyle \frac{-e^{\eta-e^\eta}+2e^\eta+3e^{-e^\eta}-3}{6(1-e^{-e^\eta})}$ \bigskip \\
     Gamma & $\displaystyle \frac{1}{\mu}$ & $\displaystyle \eta - \frac{y-\mu}{\mu^2}$ & $\displaystyle -\frac{h\eta\phi}{m}$ & $\displaystyle \frac{2}{3\eta}$ \medskip  \\
                     & $\displaystyle \log\mu$ & $\displaystyle \eta + \frac{y-\mu}{\mu}$ & $\displaystyle\frac{h\phi}{2m\eta e^{2\eta}}$ & $\displaystyle -\frac{1}{6}$ \bigskip \\
     Poisson & $\sqrt{\mu}$ & $\displaystyle \eta +\frac{y-\mu}{2\eta}$ & $\displaystyle \frac{h\eta}{2m}$ & $\displaystyle \frac{3}{2\eta}$ \medskip \\
                     & $\displaystyle \log\mu$ & $\displaystyle \eta + \frac{y-\mu}{\mu}$ & $\displaystyle \frac{h}{2me^\eta}$ & $\displaystyle -\frac{1}{3}$ \\
    \bottomrule
  \end{tabular}}
\end{center}
\label{adjusted_variates}
\end{table}

A similar calculation to that in Section~\ref{explicit} can be used to show that (\ref{mean_qfs}) can be written in terms of an IWLS step for $\bbeta$ and an appropriate update for $\phi$. In particular,
\begin{align}
  \bbeta^{(j+1)} & \leftarrow \left(\bX^\top \bW^{(j)} \bX\right)^{-1} \bX^\top \bW^{(j)}
                  \left(\bz^{(j)} + \phi^{(j)}\bxi^{(j)}  \right) \,,\notag \\
    \label{adjusted_iwls_mean_br}
  \phi^{(j+1)} & \leftarrow \phi^{(j)} \left\{1 + \phi^{(j)}\frac{\sum\left(q_i^{(j)} - \rho_i^{(j)}\right)}{\sum m_i^2
  a_i''^{(j)}} + \phi^{(j)}\frac{\sum m_i^3 a_i'''^{(j)}}{\left(\sum m_i^2
  a_i''^{(j)}\right)^2} + \left(\phi^{(j)}\right)^2 \frac{p -
  2}{\sum m_i^2a_i''^{(j)}} \right\}\,.
\end{align}
Expression~\ref{adjusted_iwls_mean_br} makes apparent that, in contrast to ML, solving the mean-bias reducing adjusted score functions in GLMs with unknown dispersion parameter involves updating $\bbeta$ and $\phi$ simultaneously. This is because $b_{\bbeta}$ generally depends on $\phi$.

Despite that the stationary point of the iterative scheme (\ref{adjusted_iwls_mean_br}) is the mean BR estimates, there is no theoretical guarantee for its convergence for general GLMs. However, substantial empirical studies have shown no evidence of divergence, even in cases in which standard IWLS (\ref{iwls}) fails to converge. Some of those empirical studies are presented in Section~\ref{sec:br_mixed}, Section~\ref{sec:br_illustrations} and Section~\ref{sec:multinomial} of the present paper.

\subsection{Median bias-reducing adjusted score functions}
\label{medianadjustedscores}

\citet{kenne+salvan+sartori:2017} introduce a family of adjusted score functions whose solution has smaller median bias than the ML estimator. Specifically, the solution $\bgamma^\dagger$ of $\bs_{\bgamma} + \bA^\dagger_{\bgamma} = 0$ is such that each of its components has probability $1/2$ of underestimating the corresponding component of the parameter $\bgamma$ with an error of order $O(n^{-3/2})$, as opposed to the error of order $O(n^{-1/2})$ for $\hat\bgamma$. A useful property of the method is that it is invariant under component-wise monotone reparameterizations in terms of the improved median bias properties of the resulting estimators.

Some tedious but straightforward algebra starting from \citet[expression~(10)]{kenne+salvan+sartori:2017}, gives that the median bias-reducing adjustments $\bA^\dagger_{\bbeta}$ and $A^\dagger_\phi$ for GLMs have the form
\begin{equation}
  \label{median_adj}
 \bA^\dagger_{\bbeta} =  \bX^\top \bW (\bxi + \bX \bu) \quad \text{and} \quad A^\dagger_\phi = \frac{p}{2\phi}+\frac{ \sum_{i = 1}^n m_i^3 a'''_i}{6\phi^2\sum_{i = 1}^n m_i^2 a''_i} \, ,
\end{equation}
where $\bu = (u_1, \ldots, u_p)^\top$ with
\begin{align}
  \label{median_u}
	u_j = [(\bX^\top \bW \bX)^{-1}]_{j}^\top \bX^\top \left[
	\begin{array}{c}
	\tilde{h}_{j,1} \left\{d_1 v'_1 / (6 v_1) - d'_1/(2 d_1)\right\} \\
	\vdots \\
	\tilde{h}_{j,n} \left\{d_n v'_n / (6 v_n) - d'_n/(2 d_n)\right\}
	\end{array}
	\right] \, .
\end{align}
In the above expressions $[\bB]_j$ denotes the $j$th row of matrix $\bB$ as a column vector, $v'_i = V'(\mu_i)$, and $\tilde{h}_{j,i}$ is the $i$th diagonal element of $\bX \bK_j \bX^T \bW$, with
\[
  \bK_j = [(\bX^\top \bW \bX)^{-1}]_{j} [(\bX^\top \bW \bX)^{-1}]_{j}^\top / [(\bX^\top \bW \bX)^{-1}]_{jj}\,,
\]
and where $[B]_{jj}$ denotes the $(j,j)th$ element of a generic matrix $B$.

Similarly to the case of mean BR, the median bias-reducing adjusted score equations can be solved using quasi-Fisher scoring or equivalently IWLS, where at the $j$th iteration
\begin{align}
  \bbeta^{(j+1)} & \leftarrow \left(\bX^\top \bW^{(j)} \bX\right)^{-1} \bX^\top \bW^{(j)}
                  \left(\bz^{(j)} + \phi^{(j)}\bxi^{(j)} \right) + \phi^{(j)}\bu^{(j)}\,, \notag \\
  \label{adjusted_iwls_median_br}
  \phi^{(j+1)} & \leftarrow \phi^{(j)} \left\{1 + \phi^{(j)}\frac{\sum\left(q_i^{(j)} - \rho_i^{(j)}\right)}{\sum m_i^2
  a_i''^{(j)}} + \phi^{(j)}\frac{\sum m_i^3 a_i'''^{(j)}}{3\left(\sum m_i^2
  a_i''^{(j)}\right)^2} + \left(\phi^{(j)}\right)^2 \frac{p}{\sum m_i^2a_i''^{(j)}} \right\}\,.
\end{align}

\begin{figure}[t]
  \caption{Illustration of the IWLS update for computing the iterates of $\bbeta$ for a given $\phi$ when performing mean BR and median BR . All quantities in the figure should be understood as being pre-multiplied by $\bW^{1/2}$. The left figure shows the addition of $\phi\bxi$ to the maximum likelihood working variates $\bz$, and the subsequent projection onto $\mathcal{C}$ (the column space of $\bW^{1/2} \bX$) that gives the updated value for the mean BR estimates $\bbeta^*$. The right figure illustrates the addition of $\phi \bu$ on $\bbeta^*$ to give the updated value for the median BR estimates $\bbeta^\dagger$.}
  \begin{center}
    \ifx\du\undefined
  \newlength{\du}
\fi
\setlength{\du}{6\unitlength}
\begin{tikzpicture}
\pgftransformxscale{1.000000}
\pgftransformyscale{-1.000000}
\definecolor{dialinecolor}{rgb}{0.000000, 0.000000, 0.000000}
\pgfsetstrokecolor{dialinecolor}
\definecolor{dialinecolor}{rgb}{1.000000, 1.000000, 1.000000}
\pgfsetfillcolor{dialinecolor}
\pgfsetlinewidth{0.100000\du}
\pgfsetdash{}{0pt}
\pgfsetdash{}{0pt}
\pgfsetmiterjoin
\pgfsetbuttcap
\definecolor{dialinecolor}{rgb}{0.898039, 0.898039, 0.898039}
\pgfsetstrokecolor{dialinecolor}
\draw (2.000000\du,29.000000\du)--(10.000000\du,24.000000\du)--(31.000000\du,24.000000\du)--(38.000000\du,29.000000\du)--cycle;
\pgfsetlinewidth{0.100000\du}
\pgfsetdash{}{0pt}
\pgfsetdash{}{0pt}
\pgfsetmiterjoin
\pgfsetbuttcap
\definecolor{dialinecolor}{rgb}{0.898039, 0.898039, 0.898039}
\pgfsetstrokecolor{dialinecolor}
\draw (40.000000\du,29.000000\du)--(48.000000\du,24.000000\du)--(69.000000\du,24.000000\du)--(76.000000\du,29.000000\du)--cycle;
\pgfsetlinewidth{0.100000\du}
\pgfsetdash{}{0pt}
\pgfsetdash{}{0pt}
\pgfsetbuttcap
{
\definecolor{dialinecolor}{rgb}{0.000000, 0.000000, 0.000000}
\pgfsetfillcolor{dialinecolor}
\pgfsetarrowsend{stealth}
\definecolor{dialinecolor}{rgb}{0.000000, 0.000000, 0.000000}
\pgfsetstrokecolor{dialinecolor}
\draw (13.000000\du,25.000000\du)--(22.000000\du,12.000000\du);
}
\definecolor{dialinecolor}{rgb}{0.000000, 0.000000, 0.000000}
\pgfsetstrokecolor{dialinecolor}
\node[anchor=west] at (30.600000\du,10.050000\du){};
\definecolor{dialinecolor}{rgb}{0.000000, 0.000000, 0.000000}
\pgfsetstrokecolor{dialinecolor}
\node[anchor=west] at (30.600000\du,9.050000\du){};
\definecolor{dialinecolor}{rgb}{0.000000, 0.000000, 0.000000}
\pgfsetstrokecolor{dialinecolor}
\node[anchor=west] at (20.000000\du,26.500000\du){};
\definecolor{dialinecolor}{rgb}{0.000000, 0.000000, 0.000000}
\pgfsetstrokecolor{dialinecolor}
\node[anchor=west] at (19.000000\du,21.000000\du){$\phi \bxi$};
\definecolor{dialinecolor}{rgb}{0.000000, 0.000000, 0.000000}
\pgfsetstrokecolor{dialinecolor}
\node[anchor=west] at (20.000000\du,23.000000\du){};
\definecolor{dialinecolor}{rgb}{0.000000, 0.000000, 0.000000}
\pgfsetstrokecolor{dialinecolor}
\node[anchor=west] at (29.600000\du,10.050000\du){};
\pgfsetlinewidth{0.100000\du}
\pgfsetdash{}{0pt}
\pgfsetdash{}{0pt}
\pgfsetbuttcap
{
\definecolor{dialinecolor}{rgb}{0.000000, 0.000000, 0.000000}
\pgfsetfillcolor{dialinecolor}
\pgfsetarrowsend{stealth}
\definecolor{dialinecolor}{rgb}{0.000000, 0.000000, 0.000000}
\pgfsetstrokecolor{dialinecolor}
\draw (13.000000\du,25.000000\du)--(28.000000\du,25.000000\du);
}
\pgfsetlinewidth{0.100000\du}
\pgfsetdash{{1.000000\du}{1.000000\du}}{0\du}
\pgfsetdash{{0.750000\du}{0.750000\du}}{0\du}
\pgfsetbuttcap
{
\definecolor{dialinecolor}{rgb}{0.000000, 0.000000, 0.000000}
\pgfsetfillcolor{dialinecolor}
\pgfsetarrowsend{stealth}
\definecolor{dialinecolor}{rgb}{0.000000, 0.000000, 0.000000}
\pgfsetstrokecolor{dialinecolor}
\draw (13.000000\du,25.000000\du)--(19.000000\du,22.000000\du);
}
\pgfsetlinewidth{0.100000\du}
\pgfsetdash{{0.750000\du}{0.750000\du}}{0\du}
\pgfsetdash{{0.750000\du}{0.750000\du}}{0\du}
\pgfsetbuttcap
{
\definecolor{dialinecolor}{rgb}{0.000000, 0.000000, 0.000000}
\pgfsetfillcolor{dialinecolor}
\pgfsetarrowsend{stealth}
\definecolor{dialinecolor}{rgb}{0.000000, 0.000000, 0.000000}
\pgfsetstrokecolor{dialinecolor}
\draw (13.000000\du,25.000000\du)--(28.000000\du,9.000000\du);
}

\definecolor{dialinecolor}{rgb}{0.000000, 0.000000, 0.000000}
\pgfsetstrokecolor{dialinecolor}
\node[anchor=north] at (2.000000\du,29.000000\du){$\mathcal{C}$};
\definecolor{dialinecolor}{rgb}{0.000000, 0.000000, 0.000000}
\pgfsetstrokecolor{dialinecolor}
\node[anchor=north] at (40.000000\du,29.000000\du){$\mathcal{C}$};

\definecolor{dialinecolor}{rgb}{0.000000, 0.000000, 0.000000}
\pgfsetstrokecolor{dialinecolor}
\node[anchor=east] at (22.000000\du,11.000000\du){$\bz$};
\definecolor{dialinecolor}{rgb}{0.000000, 0.000000, 0.000000}
\pgfsetstrokecolor{dialinecolor}
\node[anchor=west] at (28.000000\du,8.000000\du){$\bz + \phi\bxi$};
\definecolor{dialinecolor}{rgb}{0.000000, 0.000000, 0.000000}
\pgfsetstrokecolor{dialinecolor}
\node[anchor=west] at (22.000000\du,11.000000\du){};
\definecolor{dialinecolor}{rgb}{0.000000, 0.000000, 0.000000}
\pgfsetstrokecolor{dialinecolor}
\node[anchor=west] at (22.000000\du,11.000000\du){};
\definecolor{dialinecolor}{rgb}{0.000000, 0.000000, 0.000000}
\pgfsetstrokecolor{dialinecolor}
\node[anchor=west] at (16.000000\du,27.000000\du){};
\definecolor{dialinecolor}{rgb}{0.000000, 0.000000, 0.000000}
\pgfsetstrokecolor{dialinecolor}
\node[anchor=west] at (16.000000\du,27.000000\du){};
\definecolor{dialinecolor}{rgb}{0.000000, 0.000000, 0.000000}
\pgfsetstrokecolor{dialinecolor}
\node[anchor=west] at (58.000000\du,26.500000\du){};
\definecolor{dialinecolor}{rgb}{0.000000, 0.000000, 0.000000}
\pgfsetstrokecolor{dialinecolor}
\node[anchor=west] at (58.000000\du,23.000000\du){};
\pgfsetlinewidth{0.100000\du}
\pgfsetdash{}{0pt}
\pgfsetdash{}{0pt}
\pgfsetbuttcap
{
\definecolor{dialinecolor}{rgb}{0.000000, 0.000000, 0.000000}
\pgfsetfillcolor{dialinecolor}
\pgfsetarrowsend{stealth}
\definecolor{dialinecolor}{rgb}{0.000000, 0.000000, 0.000000}
\pgfsetstrokecolor{dialinecolor}
\draw (51.000000\du,25.000000\du)--(66.000000\du,25.000000\du);
}
\pgfsetlinewidth{0.100000\du}
\pgfsetdash{{\pgflinewidth}{0.200000\du}}{0cm}
\pgfsetdash{{\pgflinewidth}{0.200000\du}}{0cm}
\pgfsetbuttcap
{
\definecolor{dialinecolor}{rgb}{0.000000, 0.000000, 0.000000}
\pgfsetfillcolor{dialinecolor}
\pgfsetarrowsend{stealth}
\definecolor{dialinecolor}{rgb}{0.000000, 0.000000, 0.000000}
\pgfsetstrokecolor{dialinecolor}
\draw (51.000000\du,25.000000\du)--(53.000000\du,27.000000\du);
}
\pgfsetlinewidth{0.100000\du}
\pgfsetdash{{\pgflinewidth}{0.200000\du}}{0cm}
\pgfsetdash{{\pgflinewidth}{0.200000\du}}{0cm}
\pgfsetbuttcap
{
\definecolor{dialinecolor}{rgb}{0.000000, 0.000000, 0.000000}
\pgfsetfillcolor{dialinecolor}
\pgfsetarrowsend{stealth}
\definecolor{dialinecolor}{rgb}{0.000000, 0.000000, 0.000000}
\pgfsetstrokecolor{dialinecolor}
\draw (51.000000\du,25.000000\du)--(68.000000\du,27.000000\du);
}
\definecolor{dialinecolor}{rgb}{0.000000, 0.000000, 0.000000}
\pgfsetstrokecolor{dialinecolor}
\node[anchor=east] at (52.000000\du,27.000000\du){$\phi \bX \bu$};
\definecolor{dialinecolor}{rgb}{0.000000, 0.000000, 0.000000}
\pgfsetstrokecolor{dialinecolor}
\node[anchor=west] at (69.000000\du,28.000000\du){$\bX \bbeta^\dagger$};
\definecolor{dialinecolor}{rgb}{0.000000, 0.000000, 0.000000}
\pgfsetstrokecolor{dialinecolor}
\node[anchor=west] at (54.000000\du,27.000000\du){};
\definecolor{dialinecolor}{rgb}{0.000000, 0.000000, 0.000000}
\pgfsetstrokecolor{dialinecolor}
\node[anchor=west] at (54.000000\du,27.000000\du){};
\definecolor{dialinecolor}{rgb}{0.000000, 0.000000, 0.000000}
\pgfsetstrokecolor{dialinecolor}
\node[anchor=west] at (54.000000\du,27.000000\du){};
\definecolor{dialinecolor}{rgb}{0.000000, 0.000000, 0.000000}
\pgfsetstrokecolor{dialinecolor}
\node[anchor=west] at (27.000000\du,23.000000\du){$\bX \bbeta^*$};
\definecolor{dialinecolor}{rgb}{0.000000, 0.000000, 0.000000}
\pgfsetstrokecolor{dialinecolor}
\node[anchor=west] at (65.000000\du,23.000000\du){$\bX \bbeta^*$};
\end{tikzpicture}
  \end{center}
  \label{iwls_br_figure}
\end{figure}
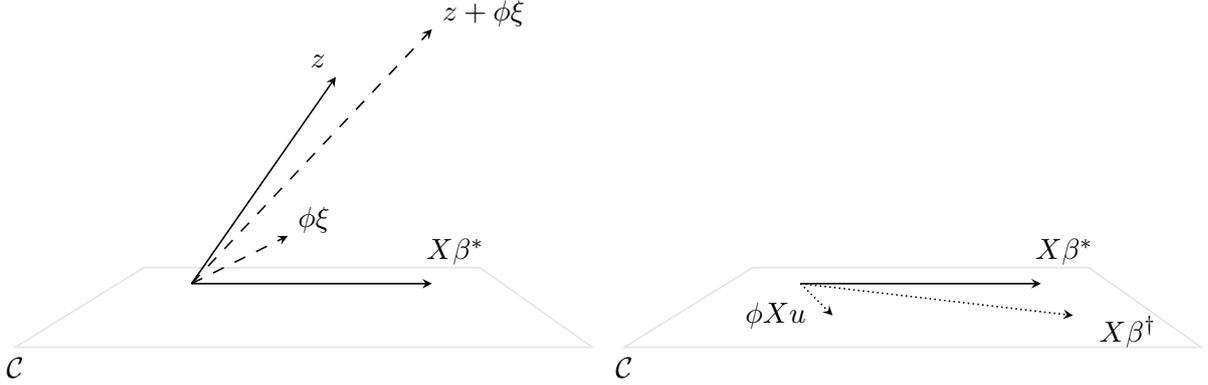

Note here that the working variate for median BR is the one for mean BR plus the extra term $\phi \bX \bu$. Equivalently, and since the extra term is in the column space of $\bX$, the median BR IWLS update for $\bbeta$ consists of a mean BR update for $\bbeta$ as in (\ref{adjusted_iwls_mean_br}), and a translation of the result by $\phi \bu$. Figure~\ref{iwls_br_figure} illustrates that procedure. The core quantities in the definition of $u$ are $d_i v'_i / (6 v_i) - d'_i/(2 d_i)$ in expression (\ref{median_u}), and Table~\ref{adjusted_variates} includes their expressions for some well-used GLMs.

Similarly to (\ref{adjusted_iwls_mean_br}), there is no theoretical guarantee for the convergence of the iterative scheme (\ref{adjusted_iwls_median_br}) for general GLMs. However, even in this case, our extensive empirical studies have produced no evidence of divergence.

\section{Inference with mean and median bias reduction}
\label{sec:br_inference}
\subsection{Wald-type inference by plug-in}

According to the results in \citet{firth:1993} and \citet{kenne+salvan+sartori:2017}, both $\btheta^*$ and $\btheta^\dagger$ have the same asymptotic distribution as the ML estimator, and hence are all asymptotically unbiased and efficient. Hence, the distribution of those estimators for finite samples can be approximated by a normal with mean $\btheta$ and variance-covariance matrix $\{\bi(\btheta)\}^{-1}$, where $\bi(\btheta)$ is given in~(\ref{information}). The derivation of this result relies on the fact that both $\bA_{\btheta}^*$ and $\bA_{\btheta}^\dagger$ are of order $O(1)$, and hence dominated by the score function as information increases.

The implication of the above results is that standard errors for the components of $\btheta^*$ and $\btheta^\dagger$ can be computed as for the ML estimator, using the square roots of the diagonal elements of $\{\bi(\bbeta^*, \phi^*)\}^{-1}$ and $\{\bi(\bbeta^\dagger, \phi^\dagger)\}^{-1}$, respectively. As a result, first-order inference, like standard Wald tests and Wald-type confidence intervals and regions are constructed in a plug-in fashion, by replacing the ML estimates with the mean BR or median BR estimates in the usual procedures in standard software.

Of course, for finite samples, Wald type procedures based on the use of ML, mean and median bias reduction will yield different results. Such differences will disappear as the samples size increases. Subsection~\ref{normal} explores those differences in normal linear regression models.

\subsection{Normal linear regression models}
\label{normal}

Consider a normal regression model with $y_1,\ldots, y_n$ realizations of independent random variables $Y_1, \ldots, Y_n$ where $Y_i$ has a $N(\mu_i,\phi/m_i)$ $(i=1,\ldots,n)$ with $\mu_i=\eta_i= \sum_{t=1}^p \beta_t x_{it}$. The adjustment terms $\bA_{\bbeta}^*$ and $\bA_{\bbeta}^\dagger$ are zero for this model. As a result, the ML, mean BR and median BR estimators of $\bbeta$ coincide with the least squares estimator $(\bX^{\top}\bM \bX)^{-1}\bX^{\top}\bM\by$, where $\bM = {\rm diag}\left\{m_1, \ldots, m_n\right\}$.  On the other hand, the ML, mean BR and median BR estimators for $\phi$ are $\hat\phi=\sum_{i=1}^n (y_i-\hat\mu_i)^2/n$, $\phi^*=\sum_{i=1}^n (y_i-\hat\mu_i)^2/(n-p)$ and $\phi^\dagger=\sum_{i=1}^n (y_i-\hat\mu_i)^2/(n-p-2/3)$.

The estimator $\phi^*$ is mean unbiased for $\phi$ and for this reason it is the default choice for estimating the precision parameter in normal linear regression models.  On the other hand, and as shown by
Theorem~\ref{the:closer} below, the use of $\phi^\dagger$ for Wald-type inference about $\beta_j$ based on asymptotic Normality, leads to inferences that are closer to the exact ones, based on the Student $t_{n-p}$ distribution, than when $\phi^*$ is used, for all practically relevant values of $n - p$ and $\alpha$.

Let $\hat{I}_{1-\alpha}=\{\hat\beta_j \pm z_{1-\alpha/2}\, (\kappa_{j}\, \hat\phi)^{1/2}\}$, ${I}_{1-\alpha}^*=\{\hat\beta_j \pm z_{1-\alpha/2}\, (\kappa_{j} \,\phi^*)^{1/2}\}$ and ${I}_{1-\alpha}^\dagger=\{\hat\beta_j \pm z_{1-\alpha/2}\, (\kappa_{j}\, \phi^\dagger)^{1/2}\}$ be the Wald-type confidence intervals for $\beta_j$ of nominal level $1-\alpha$, based on the asymptotic normal distribution of $\hat\bbeta$, $\bbeta^*$ and $\bbeta^\dagger$, respectively, where $z_\alpha$ is the quantile of level $\alpha$ of the standard normal and $\kappa_j = [(\bX^\top \bM \bX)^{-1}]_{jj}$. Let also $I^E_{1-\alpha}=\{\hat\beta_j \pm t_{n-p;1-\alpha/2} \,(\kappa_{j}\, \phi^*)^{1/2}\}$ be the confidence interval of exact level $1-\alpha$ for $\beta_j$, where $t_{n - p;\alpha}$ is the quantile of level $\alpha$ of the Student $t$ distribution with $n - p$ degrees of freedom, and define $\mathrm{Len}(I)$ to be the length of interval
$I$.

\begin{theorem}
  \label{the:closer}
For $n-p \ge 1$ and $\alpha \in (0,1)$, $\hat{I}_{1-\alpha} \subset I_{1-\alpha}^* \subset  {I}_{1-\alpha}^E$ and $I_{1-\alpha}^* \subset I_{1-\alpha}^\dagger$. 
Moreover, for $n-p \ge 1$ and $0< \alpha < 0.35562$, $
I_{1-\alpha}^\dagger \subset {I}_{1-\alpha}^E$.\\
Finally, for $n - p >1$ and $\alpha \in (0,1)$
\[
  \left|\mathrm{Len}(I_{1-\alpha}^\dagger) - \mathrm{Len}(I_{1-\alpha}^E)\right| < \left|\mathrm{Len}(I_{1-\alpha}^*) - \mathrm{Len}(I_{1-\alpha}^E)\right| \, .
\]
If $n - p =1 $, the latter inequality holds for any $0< \alpha < 0.62647$.
\end{theorem}
The proof of Theorem~\ref{the:closer} is in the Appendix.

Exact inferential solutions are not generally available for other GLMs with unknown dispersion parameter. It is therefore of interest to inverstigate whether the desirable behaviour of inference based on the median BR estimator, as demonstrated in Theorem~\ref{the:closer} for the normal linear regression model, is preserved, at least approximately, in other models. Section \ref{gamma} considers an example with Gamma regression.


\section{Mixed adjustments for dispersion models}
\label{sec:br_mixed}
\label{mixed}

In contrast to ML, mean BR is inherently not invariant to general transformations of the model parameters, in terms of its smaller asymptotic mean bias properties. This imposes a level of arbitrariness when carrying out inference on $\bbeta$ in GLMs with unknown dispersion parameters, mainly because $\phi$ appears as a factor on the variance-covariance matrix $\{\bi(\bbeta, \phi)\}^{-1}$ of the estimators. For example, standard errors for $\bbeta^*$ will be different if the bias is reduced for $\phi$ or $1/\phi$. The mean BR estimates are exactly invariant under general affine transformations, which is useful in regressions that involve categorical covariates where invariance under parameter contrasts is, typically, required. On the other hand, median BR is invariant, in terms of smaller asymptotic median bias, under componentwise monotone transformations of the parameters, but it is not invariant under more general parameter transformations, like parameter contrasts.

In order to best exploit the invariance properties of each method, we propose the default use of a mixed adjustment that combines the mean bias-reducing adjusted score for $\bbeta$ with the median bias-reducing adjusted score for $\phi$ by jointly solving
\[
\bs_{\bbeta} + \bA_{\bbeta}^* = 0_p \quad \text{and} \quad s_\phi + A_\phi^\dagger = 0\,.
\]
with $\bA_{\bbeta}^*$ and $A_\phi^\dagger$ as in expressions~(\ref{mean_adjustments}) and (\ref{median_adj}), respectively. For GLMs with known $\phi$, like Poisson or Binomial models, the mixed adjustment results in mean BR. On the contrary, for the normal linear models of Section~\ref{normal} the mixed adjustment results in median BR because $A^*_\beta = A^\dagger_\beta=0_p$.

For general GLMs with unknown $\phi$, the mixed adjustment provides the estimators $\bbeta^\ddagger$ and $\phi^\ddagger$, which are asymptotically equivalent to third order to $\bbeta^*$ and $\phi^\dagger$, respectively. The proof of this result is a direct consequence of the orthogonality \citep{cox+reid:1987} between $\bbeta$ and $\phi$ and makes use of the expansions in the Appendix of \citet{kenne+salvan+sartori:2017}. Specifically, parameter orthogonality implies that terms up to order $O(n^{-1})$ in the expansion of $\bbeta^\ddagger-\bbeta$ are not affected by terms of order $O(1)$ in $s_\phi + A^\dagger_\phi$. As a result, and up to order $O(n^{-1})$, the expansion of $\bbeta^\ddagger-\bbeta$ is the same as that of $\bbeta^* - \bbeta$.  The same reasoning applies if we switch the roles of $\bbeta$ and $\phi$, i.e. the expansion of $\phi^\ddagger-\phi$ is the same to the expansion of $\phi^\dagger-\phi$, up to order $O(n^{-1})$. Hence, $\bbeta^\ddagger$ has the same mean bias properties as $\bbeta^*$ and $\phi^\ddagger$ has the same median bias properties as $\phi^\dagger$. 
For this reason we use the term mixed BR to refer to the solution of adjusted score functions resulting from the mixed adjustment.

In order to illustrate the stated invariance properties of the estimators coming from the mixed adjustment, we consider a gamma regression model with independent response random variables $Y_1, \ldots, Y_{12}$, where, conditionally on covariates $s_{i}$ and $t_i$, each $Y_i$ has a gamma distribution with mean $\mu_i = \exp(\eta_i)$ and variance $\phi \mu_i^2$. The predictor $\eta_i$ is a function of regression parameters and the covariates, $s_i$ is a categorical covariate with values $L1$, $L2$ and $L3$, and $t_1, \ldots, t_{12}$ are generated from an exponential distribution with rate $1$. Consider the three alternative parameterizations in Table~\ref{param}. The identities $\beta_1 = \gamma_1$, $\beta_2 = \gamma_1 + \gamma_2$ and $\beta_3 = \gamma_1 + \gamma_3$ follow directly.

We simulate $1000$ independent response vectors from the parameter value $(\beta_1, \beta_2, \beta_3, \beta_4, \phi)^\top$ $=$ $(-1, -0.5, 3, 0.2, 0.5)^\top$, and estimate the three parameter vectors in Table~\ref{param} for each sample using the ML estimator, and the estimators resulting from the mean, median and mixed bias-reducing adjusted scores. The estimates for parameterizations I and III are used to estimate the probability $P(|\tilde{\beta_2} - \tilde{\gamma_1} - \tilde{\gamma_2}| > \epsilon_1)$, and those for parameterizations I and II are used to estimate the probability $P(|\tilde{\phi} - \exp(\tilde{\zeta})| > \epsilon_2)$ for various values of $\epsilon_1$ and $\epsilon_2$, using the various estimators in place of $\tilde{\beta}_2$, $\tilde{\gamma}_1$, $\tilde{\gamma}_2$, $\tilde{\phi}$ and $\tilde{\zeta}$. The results are displayed in Table \ref{probs}. As expected, the probability $P(|\tilde{\beta_2} - \tilde{\gamma_1} - \tilde{\gamma_2}| > \epsilon_1)$ is zero for ML and mean BR, but not for median BR. Similarly, the probabbility $P(|\tilde{\phi} - \exp(\tilde{\zeta})| > \epsilon_2)$ is zero for ML and median BR, but not for mean BR. In contrast, the mixed adjustment strategy inherits the relevant properties of mean and median BR, and delivers estimators that are numerically invariant under linear contrasts of the mean regression parameters, and monotone transformations of the dispersion parameter.

\begin{table}[t]
  \caption{Alternative, equivalent parameterizations of a gamma regression model with independent responses $Y_1, \ldots, Y_{12}$ where, conditionally on covariates, each $Y_i$ has a gamma distribution with mean $\mu_i = \exp(\eta_i)$ and variance $\phi \mu_i^2$. The covariates $x_{i1}$, $x_{i2}$ and $x_{i3}$ encode the levels of a $3$-level categorical covariate $s_i$ as follows: $x_{i1}$ is $1$ for $i = 1, 2, 3, 4$ and $0$, otherwise, $x_{i2}$ is $1$ for $i = 5, 6, 7, 8$ and $0$, otherwise, and $x_{i3}$ is $1$ for $i = 9, 10, 11, 12$ and $0$, otherwise. The covariate values $t_1, \ldots, t_{12}$ are generated from an exponential distribution with rate $1$.}
  \begin{center}
  \begin{tabular}{cccc}
    \toprule
    Parameterization & Predictor $\eta_i$ & Dispersion $\phi$ & Parameter vector \\ \midrule
    I & $\beta_1x_{i1} + \beta_2 x_{i2} + \beta_3 x_{i3} + \beta_4 t_i$ & $\phi$ & $(\beta_1, \beta_2, \beta_3, \beta_4, \phi)^\top$\\
    II & $\beta_1x_{i1} + \beta_2 x_{i2} + \beta_3 x_{i3} + \beta_4 t_i$ & $e^\zeta$ & $(\beta_1, \beta_2, \beta_3, \beta_4, \zeta)^\top$ \\
    III & $\gamma_1 + \gamma_2 x_{i2} + \gamma_3 x_{i3} + \beta_4 t_i$ & $\phi$ & $(\gamma_1, \gamma_2, \gamma_3, \beta_4, \phi)^\top$ \\ \bottomrule
  \end{tabular}
\end{center}
\label{param}
\end{table}

\begin{table}[t]
  \caption{The probability $P(|\tilde{\beta_2} - \tilde{\gamma_1} - \tilde{\gamma_2}| > \epsilon_1)$ for parameterizations I and III, and $P(|\tilde{\phi} - \exp(\tilde{\zeta})| > \epsilon_1)$ for parameterizations I and II for various values of $\epsilon$. The ML estimator, the estimators from the mean, median and mixed bias-reducing adjusted scores are used in place of the tilded quantities. The figures are based on $1000$ simulated response   vectors from the gamma regression model of Table~\ref{param} with $(\beta_1, \beta_2, \beta_3, \beta_4, \phi)^\top = (-1, -0.5, 3, 0.2, 0.5)^\top$.}
\begin{center}
\begin{tabular}{ccccccccccc}
  \toprule
  $\epsilon_1$ & \multicolumn{4}{c}{$P(|\tilde{\beta_2} - \tilde{\gamma_1} - \tilde{\gamma_2}| > \epsilon_1)$} & & $\epsilon_2$ & \multicolumn{4}{c}{$P(|\tilde{\phi} - \exp(\tilde{\zeta})| > \epsilon_2)$} \\ \cmidrule{1-5} \cmidrule{7-11}
 & ML & \parbox{1.3cm}{\centering mean BR} & \parbox{1.3cm}{\centering median BR} & \parbox{1.3cm}{\centering mixed BR} & & & ML & \parbox{1.3cm}{\centering mean BR} & \parbox{1.3cm}{\centering median BR} & \parbox{1.3cm}{\centering mixed BR} \\ \cmidrule{2-5} \cmidrule{8-11}
0.01 & 0 & 0 & 0.656 & 0 & & 0.02 & 0 & 0.978 & 0 & 0 \\
0.02 & 0 & 0 & 0.162 & 0 & & 0.04 & 0 & 0.771 & 0 & 0 \\
0.03 & 0 & 0 & 0.034 & 0 & & 0.06 & 0 & 0.454 & 0 & 0 \\
0.04 & 0 & 0 & 0.010 & 0 & & 0.08 & 0 & 0.181 & 0 & 0 \\
0.05 & 0 & 0 & 0.003 & 0 & & 0.10 & 0 & 0.061 & 0 & 0 \\  \bottomrule
\end{tabular}
\end{center}
\label{probs}
\end{table}

Section~\ref{gamma} further evaluates the use of the mixed adjustment in the
estimation of Gamma regression models.



\section{Illustrations and simulation studies}
\label{sec:br_illustrations}
\subsection{Case studies and simulation experiments}
In this section, we present results from case-studies and confirmatory simulation studies that provide empirical support to the ability of mean and median BR to achieve their corresponding goals, i.e. mean and median bias reduction, respectively. In particular, in Section \ref{gamma} we consider gamma regression, in which we also evaluate the mixed adjustment strategy of Section \ref{mixed}, while in Section \ref{sec:birthweight} we consider logistic regression, showing how both mean and median BR provide a practical solution to the occurrence of infinite ML estimates. Finally, Section \ref{sec:infertility} evaluates the performance of mean and median BR in a logistic regression setting characterized by the presence of many nusiance parameters. In this case, ML estimation and inference are known to be unreliable, while both mean and median BR practically reproduce the behaviour of estimation and inference based on the conditional likelihood, which, in this particular case, is the gold standard.

All numerical computations are performed in R using the \texttt{brglm2} R package \citep{brglm2}. The \texttt{brglm2} R package provides the \texttt{brglmFit} method for the \texttt{glm} R function that implements mean and median BR for any GLM using the quasi-Fisher scoring iteration introduced in Section~\ref{sec:br_iwsm}.

\subsection{Gamma regression model for blood clotting times}
\label{gamma}

\begin{table}[t]
  \caption{Clotting data. Estimates and estimated standard errors (in parenthesis) for the parameters of the model in Example~\ref{clotting}.}
\label{estimategammamodel}
\begin{center}
\begin{tabular}{lrrrrr}
\toprule
Method & \multicolumn{1}{c}{$\beta_1$} & \multicolumn{1}{c}{$\beta_2$} & \multicolumn{1}{c}{$\beta_3$} & \multicolumn{1}{c}{$\beta_4$} & \multicolumn{1}{c}{$\phi$} \\
 \midrule
ML & 5.503\phantom{)} & -0.584\phantom{)} & -0.602\phantom{)} & 0.034\phantom{)} & 0.017 \\
  & (0.161) & (0.228) & (0.047) & (0.066) & \\ \midrule
 mean BR & 5.507\phantom{)} & -0.584\phantom{)} & -0.602\phantom{)} & 0.034\phantom{)} & 0.022 \\
  & (0.183) & (0.258) & (0.053) & (0.075) & \\ \midrule
 median BR & 5.505\phantom{)} & -0.584\phantom{)} & -0.602\phantom{)} & 0.034\phantom{)} & 0.024 \\
  & (0.187) & (0.265) & (0.054) & (0.077) & \\ \midrule
 mixed BR & 5.507\phantom{)} & -0.584\phantom{)} & -0.602\phantom{)} & 0.034\phantom{)} & 0.024 \\
  & (0.187) & (0.265) & (0.054) & (0.077) & \\
  \bottomrule
\end{tabular}
\end{center}
\end{table}

The regression model for the clotting data in Example \ref{clotting} is fitted, here, using the mean, median and mixed bias-reducing adjusted score functions of Section~\ref{meanadjustedscores}, Section~\ref{medianadjustedscores} and Section~\ref{mixed}, respectively. The estimates and the corresponding estimated standard errors are reported in Table \ref{estimategammamodel}. The estimates of regression parameters are practically the same for all methods. More marked differences between ML and the three adjusted score methods are noted in the estimates of the dispersion parameter. In particular, the estimates from the adjusted score methods result in notable inflation of the estimated standard errors for the regression parameters, with the median and mixed bias-reducting adjustments resulting in the largest inflation.

\begin{table}[t!]
  \caption{Clotting data. Simulation results based on $10\, 000$ samples under the ML fit. The quantities in the table are described in the caption of Table \ref{tab:estimators2.1}. The estimators considered are those from mean BR (Section \ref{meanadjustedscores}), median BR (Section \ref{medianadjustedscores}) and mixed BR (Section \ref{mixed}). All reported figures are $\times 100$ of their actual value and $<0.01$ is used for a value that is less than $0.01$ in absolute value.}
    \begin{center}
    \begin{tabular}{lcrrrrrr}
      \toprule
    Method &\multicolumn{1}{c}{Parameter} & \multicolumn{1}{c}{B} &
                                                               \multicolumn{1}{c}{RMSE}
      & \multicolumn{1}{c}{$\text{B}^2/\text{SD}^2$} &
                                                       \multicolumn{1}{c}{PU}
      & \multicolumn{1}{c}{MAE} & \multicolumn{1}{c}{C} \\ \midrule

  {mean BR} & $\beta_1$ & -0.04 & 16.15 & $<$0.01 & 49.65 & 12.87 & 93.12 \\
  & $\beta_2$ & 0.36 & 23.09 & 0.02 & 49.59 & 18.46 & 92.69 \\
  & $\beta_3$ & 0.02 & 4.69 & $<$0.01 & 49.92 & 3.74 & 93.08 \\
 & $\beta_4$ & -0.11 & 6.71 & 0.03 & 50.50 & 5.36 & 92.26 \\
  & $\phi$ & $<$0.01 & 0.67 & $<$0.01 & 55.00 & 0.53 & \\
 \midrule
  {median BR} &$\beta_1$ & -0.15 & 16.15 & 0.01 & 49.93 & 12.87 & 93.67 \\
  & $\beta_2$ & 0.36 & 23.09 & 0.02 & 49.60 & 18.46 & 93.27 \\
  & $\beta_3$ & 0.03 & 4.69 & 0.01 & 49.88 & 3.74 & 93.73 \\
  & $\beta_4$ & -0.11 & 6.71 & 0.03 & 50.50 & 5.36 & 93.05 \\
  & $\phi$ & 0.09 & 0.71 & 1.67 & 49.99 & 0.55 & \\
   \midrule
  {mixed} & $\beta_1$ & -0.02 & 16.15 & $<$0.01 & 49.65 & 12.87 & 93.66 \\
  & $\beta_2$ & 0.36 & 23.09 & 0.02 & 49.59 & 18.46 & 93.28 \\
  & $\beta_3$ & 0.02 & 4.69 & $<$0.01 & 49.95 & 3.74 & 93.71  \\
  & $\beta_4$& -0.11 & 6.71 & 0.03 & 50.50 & 5.36 & 93.06 \\
  & $\phi$ & 0.09 & 0.71 & 1.68 & 49.93 & 0.55 & \\
               \bottomrule
    \end{tabular}
  \end{center}
  \label{tab:estimators2}
\end{table}

In order to assess the quality of the estimates in Table~\ref{estimategammamodel}, the simulated data sets in Example~\ref{clotting} are used to estimate the bias, the root mean squared error, the percentage of underestimation, and the mean absolute error of the various estimators, and the coverage of nominally $95$\% Wald-type confidence intervals. Table~\ref{tab:estimators2} reports the results. A comparison with the results for ML in Table~\ref{tab:estimators2.1} shows that the ML, mean BR, median BR and mixed BR estimators of $\beta_1, \ldots, \beta_4$ have similar bias and variance properties. On the other hand, the mean BR estimator of the dispersion parameter almost fully compensates for the mean bias of the ML estimator, while median BR and mixed BR give almost exactly $50\%$ probability of underestimation. Furthermore, all BR methods deliver marked improvements in terms of empirical coverage over ML, and the confidence intervals based on the estimates from the median and mixed bias-reducing adjustments are behaving the best. Finally, all confidence intervals appear to be liberal in terms of coverage, most probably due to the small sample size and the need to estimate the dispersion parameter.
Note here that the superior coverage when using estimates from median and mixed bias-reduction adjustments of the scores are similar to what is expected in the case of the normal linear model; see Section~\ref{normal}.

\subsection{Logistic regression for infant birth weights}
\label{sec:birthweight}
We consider a study of low birth weight using the data given in \citet[Table 2.1]{hosmer+Lemeshow:2000}, which are also publicly available in the {\tt MASS} R package. The focus here is on the 100 births for which the mother required no physician visits during the first trimester. The outcome of interest is a proxy of infant birth weight ($1$ if $\geq 2500g$ and $0$ otherwise), whose expected value $\mu_i$ is modelled in terms of explanatory variables using a logistic regression model with $\log\{\mu_i/(1-\mu_i)\}=\sum_{t=1}^7 \beta_t x_{it}$, where $x_{i1}=1$, $x_{i2}$ and $x_{i3}$ are the age and race (1 if white, 0 otherwise) of the mother, respectively, $x_{i4}$ is the mother's smoking status during pregnancy (1 if yes, 0 if no), $x_{i5}$ is a proxy of the history of premature labor (1 if any, 0 if none), $x_{i6}$ is history of hypertension (1 if yes, 0 if no), and $x_{i7}$ is the logarithm of the mother's weight at her last menstrual period.


Table~\ref{estimatelogisticmodel} gives the parameter estimates from ML, mean BR and median BR. Both mean BR and median BR deliver estimates that are shrunken versions of the corresponding ML estimates, with mean BR delivering the most shrinkage. This shrinkage translates to smaller estimated standard errors for the regression parameters. \citet{kosmidis+firth:2018} provide geometric insights for the shrinkage induced by mean BR in binary regression and prove that the mean BR estimates are always finite for full rank $X$.

The frequency properties of the resulting estimators are assessed by simulating 10\,000 samples at the ML estimates in Table~\ref{estimatelogisticmodel}, with covariates fixed as in the observed sample, and re-estimating the model from each simulated sample. A total of 103 out of the 10\,000 samples results in ML estimates with one or more infinite components due to data separation \citep{albert+anderson:1984}. The detection of infinite estimates was done prior to fitting the model using the linear programming algorithms in \citet{konis07}, as implemented in the \texttt{detect\_separation} method of the \texttt{brglm2} R package \citep{brglm2}. The separated data sets were excluded when estimating the bias and coverage of Wald-type confidence intervals for the ML estimator. In contrast, the estimates from mean and median BR estimates were finite in all cases. For this reason, the corresponding summaries are based on all 10\,000 samples.

Table \ref{logisticbiaspu} shows the results. Both mean BR and median BR have excellent performance in terms of mean bias and probability of underestimation, respectively. Table \ref{logisticbiaspu} also includes summaries for the estimators $\hat\psi_t = e^{\hat\beta_t}$ , $\psi_t^* = e^{\beta_t^*}$ , $\psi_t^\dagger = e^{\beta_t^\dagger}$ of the odds-ratios $\psi_t = e^{\beta_t}$. Estimators of $\psi_t$ with improved bias properties have also been recently investigated in \citet{lyles+guo+greenland:2012}. The invariance properties of ML and median BR guarantee that $\hat\bpsi$ and $\bpsi^\dagger$ are the ML and median BR estimators of $\bpsi$, respectively. As a result, $\psi_t^\dagger$ preserves its improved median bias properties. On the other hand, $\psi_t^*$ is not, formally, the mean BR estimator of $\bpsi$. Nevertheless, it behaves best in terms of bias. The improved estimation and inference provided by mean and median BR become even more evident in more extreme modelling settings, as shown by the example in the next section.

\begin{table}[t]
\caption{Estimates and estimated standard errors (in parenthesis) for the logistic regression model for the infant birth weight data in Section~\ref{sec:birthweight}.}
\label{estimatelogisticmodel}
\begin{center}
\begin{tabular}{lrrrrrrr}
\toprule
Method  & \multicolumn{1}{c}{$\beta_1$} &
                                          \multicolumn{1}{c}{$\beta_2$} & \multicolumn{1}{c}{$\beta_3$} & \multicolumn{1}{c}{$\beta_4$} & \multicolumn{1}{c}{$\beta_5$} & \multicolumn{1}{c}{$\beta_6$} & \multicolumn{1}{c}{$\beta_7$} \\
  \midrule
{ML} & -8.496\phantom{)} & -0.067\phantom{)} & 0.690\phantom{)} & -0.560\phantom{)} & -1.603\phantom{)} & -1.211\phantom{)} & 2.262\phantom{)} \\
   & (5.826) & (0.053) & (0.566) & (0.576) & (0.697) & (0.924) &
                                                                 (1.252) \\ \midrule
 {mean BR} & -7.401\phantom{)} & -0.061\phantom{)} & 0.622\phantom{)} & -0.531\phantom{)} & -1.446\phantom{)} & -1.104\phantom{)} & 1.998\phantom{)} \\
   & (5.664) & (0.052) & (0.552) & (0.564) & (0.680) & (0.901) &
                                                                 (1.216) \\ \midrule
{median BR} & -7.641\phantom{)} & -0.062\phantom{)} & 0.638\phantom{)} & -0.538\phantom{)} & -1.481\phantom{)} & -1.134\phantom{)} & 2.059\phantom{)} \\
   & (5.717) & (0.053) & (0.557) & (0.568) & (0.681) & (0.906) & (1.228) \\
   \bottomrule
\end{tabular}
\end{center}
\end{table}

\begin{table}[t]
  \caption{Simulation results based on $10\, 000$ samples under the ML fit of the model for the birth weight data in Subsection~\ref{sec:birthweight}. All reported summaries, described in the caption of Table \ref{tab:estimators2.1}, for ML are conditional to the finiteness of the estimates. B$_\psi$ is the estimated bias in the $\psi$ parameterization and $<0.01$ is used for a value that is less than $0.01$ in absolute value.}
  \label{logisticbiaspu}
  \begin{center}
\begin{tabular}{llrrrrrrr}
\toprule
 & Method & \multicolumn{1}{c}{$\beta_1$} &
                                            \multicolumn{1}{c}{$\beta_2$} & \multicolumn{1}{c}{$\beta_3$} & \multicolumn{1}{c}{$\beta_4$} & \multicolumn{1}{c}{$\beta_5$} & \multicolumn{1}{c}{$\beta_6$} & \multicolumn{1}{c}{$\beta_7$} \\
  \midrule
  B & ML &-1.42 & -0.01 & 0.09 & -0.03 & -0.20 & -0.12 & 0.34 \\
  & mean BR & -0.08 & $<$0.01 & 0.01 & $<$0.01 & -0.01 & $<$0.01 & 0.02 \\
  & median BR & -0.38 & $<$0.01 & 0.03 & -0.01 & -0.07 & -0.04 & 0.09 \\
  \midrule
  B$_\psi$ & ML &183.50 & $<$0.01 & 0.75 & 0.12 & 0.02 & 0.18 & 57.50 \\
  & mean BR & 47.17 & $<$0.01 & 0.41 & 0.11 & 0.05 & 0.17 & 18.75 \\
  & median BR & 56.66 & $<$0.01 & 0.50 & 0.11 & 0.04 & 0.21 & 23.74 \\
  \midrule
  RMSE & ML &  6.86 & 0.06 & 0.66 & 0.66 & 0.82 & 1.11 & 1.49 \\
  & mean BR &  5.94 & 0.05 & 0.58 & 0.59 & 0.72 & 0.94 & 1.28 \\
  & median BR & 6.11 & 0.06 & 0.60 & 0.61 & 0.78 & 1.01 & 1.32 \\
  \midrule
  PU & ML & 56.1 & 53.3 & 46.4 & 51.4 & 57.8 & 53.5 & 43.1 \\
  & mean BR & 48.2 & 49.2 & 51.3 & 49.6 & 48.1 & 48.9 & 52.2 \\
  & median BR & 50.0 & 49.6 & 49.9 & 49.9 & 50.6 & 50.3 & 50.0 \\
  \midrule
  C & ML& 94.8 & 94.8 & 94.5 & 94.7 & 96.4 & 96.6 & 94.5 \\
  & mean BR & 96.3 & 96.2 & 96.0 & 96.2 & 97.2 & 98.1 & 96.1 \\
  & median BR& 96.1 & 96.0 & 95.8 & 95.9 & 97.0 & 97.8 & 96.0 \\
  \bottomrule
\end{tabular}
\end{center}
\end{table}

\subsection{Logistic regression for the link between sterility and abortion}
\label{sec:infertility}

We consider data from a retrospective, matched case-control study on the role of induced and spontaneous abortions in the aetiology of secondary sterility \citep{trichopoulos+handanos+danezis+etal:1976}. The data are
available in the \texttt{infert} data frame from the \texttt{datasets} R package. The two healthy control subjects from the same hospital were matched to each of 83 patients according to their age, parity, and level of education. One of the cases could be matched with only one control, thus there is a total of 248 records. Each record also provides the number of induced and spontaneous abortions, taking values 0, 1 and 2 or more.

As is meaningful for retrospective case-control studies \citep[see, for example,][Section~4.3.3]{mccullagh+nelder:1989}, we consider a logistic regression model with one fixed-effect for each matched combination of cases and controls, and the number of induced and spontaneous abortions as the two categorical covariates of interest. In particular, the log-odds of secondary sterility for the $j$th individual in the $i$th case-controls combination are assumed to be
\begin{equation}
  \label{infert}
\lambda_i + \beta_1 x_{ij} + \beta_2 x'_{ij} + \beta_3 z_{ij} +
\beta_4 z'_{ij}   \quad (i = 1, \ldots, 83; j = 1, \ldots, n_i) \, ,
\end{equation}
where $n_i \in \{2, 3\}$, $x_{ij}$, $x'_{ij}$ are indicator variables of 1 and 2 or more spontaneous abortions, respectively, and $z_{ij}$ and $z'_{ij}$ are indicator variables of 1 and 2 or more induced abortions, respectively. The parameters $\lambda_1, \ldots, \lambda_{83}$ are the fixed-effects for each matched combination of cases and controls, and the parameters of interest are $\beta_1, \ldots, \beta_4$.

Due to the many nuisance parameters, the maximum likelihood estimators of $\beta_1, \ldots, \beta_4$ are highly biased leading to misleading inference. A solution that is specific to logistic regression is to eliminate the fixed-effects by conditioning on their sufficient statistics and maximize the conditional likelihood (CL). This can be done, for example, using the \texttt{clogit} function in the \texttt{survival} R package. As shown in
Table~\ref{estimatelogisticmodel2}, both mean and median BR give estimates that are close to the maximum CL estimates, practically removing all the bias from the ML estimates, and resulting also in a correction for the estimated standard errors.

This desirable behaviour of mean BR and median BR is in line with published theoretical results in stratified settings with nuisance parameters. In particular, \cite{lunardon:2018} has recently shown that inferences based on mean BR in stratified settings with strata-specific nuisance parameters are valid under the same conditions for the validity of inference \citep{sartori:2003} based on modified profile likelihoods \citep[see, e.g.][]{barndorff-nielsen:1983, cox+reid:1987, mccullagh+tibshirani:1990, severini:1998}. The same equivalence is shown for median BR in \cite{kenne+salvan+sartori:2017}.

The advantage of mean and median BR over maximum CL is their generality of application. As is shown in Table~\ref{adjusted_variates} mean and median BR can be used in models where a sufficient statistic does not exist and hence direct elimination of the nuisance parameters is not possible.  One such example is probit regression, which is typically the default choice in many econometric applications stemming out from prospective studies. The further algorithmic simplicity for mean and median BR make them also competitive to the various modified profile likelihoods.

\begin{table}[t]
  \caption{Estimates and estimated standard errors (in parenthesis) for the parameters of interest in model~(\ref{infert}) for the sterility data in Subsection~\ref{sec:infertility}.}
\label{estimatelogisticmodel2}
\begin{center}
\begin{tabular}{lrrrr}
\toprule
Method  & \multicolumn{1}{c}{$\beta_1$} &
                                          \multicolumn{1}{c}{$\beta_2$} & \multicolumn{1}{c}{$\beta_3$} & \multicolumn{1}{c}{$\beta_4$} \\
  \midrule
{ML} & 3.268 (0.592)  &  6.441 (0.955)  &  2.112 (0.587)  &  4.418 (0.948) \\
{CL} & 2.044 (0.453)  &  3.935 (0.725)  &  1.386 (0.463)  &  2.819 (0.735)  \\
{mean BR} & 2.055 (0.472) & 3.954 (0.708) & 1.305 (0.474) & 2.714 (0.744)\\
{median BR} & 2.083 (0.478) & 3.997 (0.713) & 1.330 (0.482) & 2.760 (0.754)\\
\bottomrule
\end{tabular}
\end{center}
\end{table}


\section{Multinomial logistic regression}
\label{sec:multinomial}
\subsection{The Poisson trick}

Suppose that $\by_1, \ldots, \by_n$ are $k$-vectors of counts with $\sum_{j = 1}^k y_{ij} = m_i$ and that $\bx_1, \ldots, \bx_n$ are corresponding $p$-vectors of explanatory variables. The multinomial logistic regression model assumes that conditionally on $\bx_1, \ldots, \bx_n$ the vectors of counts $\by_1, \ldots, \by_n$ are realizations of independent multinomial vectors, with $y_i=(y_{i1},\ldots,y_{ik})$, where the probabilities for the $i$th multinomial vector satisfy
\begin{equation}
  \label{baseline_logit}
  \log\frac{\pi_{ij}}{\pi_{ik}} = \bx_i^\top \bgamma_j \quad (j = 1, \ldots, k - 1) \, ,
\end{equation}
with $\sum_{j = 1}^k \pi_{ij} = 1$. Typically, $x_{i1} = 1$ for every $i \in \{1,\ldots, n\}$. The above model is also known as the baseline category logit \citep[see, for example,][\S 7.1]{agresti:2002} because it uses one of the multinomial categories as a baseline for the definition of the log-odds. Expression~(\ref{baseline_logit}) has the $k$th category as baseline, but this is without loss of generality since any other log-odds can be computed using simple contrasts of the parameter vectors $\bgamma_1, \ldots, \bgamma_{k-1}$.

Maximum likelihood estimation can be done either by direct maximization of the multinomial log-likelihood for~(\ref{baseline_logit}) or using maximum likelihood for an equivalent Poisson log-linear model. Specifically, if $y_{11}, \ldots, y_{nk}$ are realizations of independent Poisson random variables with means $\mu_{11}, \ldots, \mu_{nk}$, where
\begin{align}
  \label{poison_log}
  \log\mu_{ij} & = \lambda_i + \bx_i^\top \bgamma_j \quad (j = 1, \ldots, k -
                 1)\,, \\ \notag
  \log\mu_{ik} & = \lambda_i\, ,
\end{align}
then the score equations for $\lambda_i$ are $m_i = \sum_{j = 1}^k \mu_{ij}$, forcing the Poisson means to add up to the multinomial totals and the maximum likelihood estimates for $\bgamma_1, \ldots, \bgamma_{k-1}$ to be exactly those that result from maximising the multinomial likelihood for model~(\ref{baseline_logit}) directly.

\citet{kosmidis+firth:2011} proved that the equivalence of the multinomial logistic regression model~(\ref{baseline_logit}) and the Poisson log-linear model~(\ref{poison_log}) extends to the mean BR estimates of $\bgamma_1, \ldots, \bgamma_{k-1}$, if at each step of the iterative procedure for solving the adjusted score equations, the current values of the Poisson expectations $\mu_{i1}, \ldots, \mu_{ik}$ are rescaled to sum up to the corresponding multinomial totals. Specifically, the results in \citet{kosmidis+firth:2011} suggest to prefix the IWLS update in~(\ref{adjusted_iwls_mean_br}) for the Poisson log-linear model~(\ref{poison_log}) with the extra step
\[
  \bar{\mu}_{is}^{(j)} \leftarrow m_{is} \frac{\mu_{is}^{(j)}}{\sum_{t =
      1}^k\mu_{it}^{(j)}} \quad (i = 1,\ldots,n; s = 1,\ldots, k)
\]
that rescales the Poisson means to sum to the multinomial totals. Then, $W$ and the ML and mean BR quantities in the last row of Table~\ref{adjusted_variates} are computed using $\bar{\mu}_{is}^{(j)}$ instead of $\mu_{is}^{(j)}$.

The same argument applies the case of median BR. Given that the extra term in the IWLS update for median bias reduction in~(\ref{adjusted_iwls_median_br}) depends on the parameters only through the response means, the same extra step of rescaling the Poisson means before the IWLS update of the parameters, will result in an iteration that delivers the median BR estimates of the multinomial logistic regression model using the equivalent Poisson log-linear model.

\subsection{Invariance properties}
The mean BR estimator is invariant under general affine transformations of the parameters, and hence, direct contrasts result in mean BR estimators for any other baseline category for the response and any reference category in the covariates, without refitting the model. This is a particularly useful guarantee when modelling with baseline category models. In contrast, a direct transformation of the median BR estimates with baseline category $k$ or a specific set of contrasts for the covariates, is not guaranteed to result in median BR estimates for other baseline categories or contrasts in general.

\subsection{Primary food choices of alligators}
In order to investigate the extent that non-invariance impacts estimation and inference, we consider the data on food choice of alligators analyzed in \citet[Section 7.1.2]{agresti:2002}. The data comes from a study of factors influencing the primary food choice of alligators. The observations are 219 alligators captured in four lakes in Florida. The nominal response variable is the primary food type, in volume, found in an alligator's stomach, which has five categories (fish, invertebrate, reptile, bird, other).  The dataset classifies the primary food choice according to the lake of capture (Hancock, Oklawaha, Trafford, George), gender (male, female), and size of the alligator ($\le 2.3$ meters long, $>2.3$ meters long).

Let $s=1$ for alligator size $> 2.3$ meters and 0 otherwise, and let $z^H, z^O, z^T, z^G$ be indicator variables for the lakes; for instance, $z^H=1$ for alligators on the lake Hancock and 0 otherwise. A possible model for the probabilities of food choice is
\begin{equation}
\label{multinomialmodel2}
\log(\pi_{ic}/\pi_{i1})=\gamma_{c1} +
\gamma_{c2}s_i+\gamma_{c3}z^O_i+\gamma_{c4}z^T_i+\gamma_{c5}z^G_i \quad
(c=2,3,4,5)\,,
\end{equation}
where $\pi_{ic}$ is the probability for category $c$, with values corresponding to fish ($c = 1$), invertebrate ($c = 2$), reptile ($c = 3$), bird ($c = 4$) and other ($c = 5$). Model (\ref{multinomialmodel2}) is based on the choice of contrasts that would be selected by default in R. In order to investigate the effects of lack of invariance of median bias reduction, the set of contrasts used  in \citet[Section 7.1.2]{agresti:2002} is considered where George is the reference lake and $> 2.3$ is the reference alligator size. These choices result in writing the food choice log-odds as
\begin{equation}
\label{multinomialmodel1}
\log(\pi_{ic}/\pi_{i1})=\gamma_{c1}^{\prime} +
\gamma_{c2}^{\prime}s_i^{\prime}+\gamma_{c3}^{\prime}z^H_i+\gamma_{c4}^{\prime}z^O_i+\gamma_{j5}^{\prime}z^T_i
\quad (c=2,3,4,5),
\end{equation}
where $s^{\prime}=1$ for alligator size $\le 2.3$ meters and 0 otherwise. The coefficients in the linear predictors of (\ref{multinomialmodel2}) and (\ref{multinomialmodel1}) are related as $\gamma_{c1}=\gamma_{c1}^{\prime}+\gamma_{c2}^{\prime}+ \gamma_{c3}^{\prime}$, $\gamma_{c2}=-\gamma_{c2}^{\prime}$, $\gamma_{c3}=\gamma_{c4}^{\prime}-\gamma_{c3}^{\prime}$, $\gamma_{c4}=\gamma_{c5}^{\prime}-\gamma_{c3}^{\prime}$ and $\gamma_{c5}=-\gamma_{c3}^{\prime}$.

\begin{table}[t]
\caption{Estimates and estimated standard errors (in parenthesis) of the multinomial regression model (\ref{multinomialmodel2}) for the alligator data in Section~\ref{sec:multinomial}.}
\label{estimatemultinomialmodel}
\begin{center}
\begin{tabular}{lrrrrrr}
\toprule
Method  & \multicolumn{1}{c}{$c$} & \multicolumn{1}{c}{$\gamma_{c1}$}
  & \multicolumn{1}{c}{$\gamma_{c2}$} &
                                        \multicolumn{1}{c}{$\gamma_{c3}$} & \multicolumn{1}{c}{$\gamma_{c4}$} & \multicolumn{1}{c}{$\gamma_{c5}$}  \\
  \midrule
ML & $2$ & -1.75  (0.54) & -1.46  (0.40) & 2.60 (0.66) & 2.78 (0.67) & 1.66 (0.61) \\
  & $3$ & -2.42 (0.64) & 0.35 (0.58) & 1.22  (0.79) & 1.69  (0.78) & -1.24 (1.19) \\
  & $4$ & -2.03  (0.56) & 0.63  (0.64) & -1.35  (1.16) & 0.39  (0.78) & -0.70  (0.78) \\
  & $5$ & -0.75 (0.35) & -0.33 (0.45) & -0.82 (0.73) & 0.69 (0.56) & -0.83 (0.56) \\
  \cline{2-7}
  mean BR & $2$ & -1.65 (0.52) & -1.40 (0.40) & 2.46 (0.65) & 2.64  (0.66) & 1.56 (0.60) \\
  & $3$ & -2.25 (0.61) & 0.32 (0.56) & 1.12 (0.76) & 1.58 (0.75) & -0.98 (1.02) \\
  & $4$ & -1.90 (0.54) & 0.58 (0.61) & -1.04 (1.01) & 0.40 (0.76) & -0.62 (0.74) \\
  & $5$ & -0.72 (0.35) & -0.31 (0.44) & -0.72 (0.71) & 0.67 (0.56) & -0.78 (0.55) \\
    \cline{2-7}
  median BR & $2$ & -1.71 (0.53) & -1.41 (0.40) & 2.51 (0.65) & 2.69 (0.67) & 1.61 (0.61) \\
  & $3$ & -2.33 (0.62) & 0.34 (0.57) & 1.16 (0.77) & 1.62 (0.76) & -1.12 (1.10) \\
 & $4$& -1.96 (0.54) & 0.60 (0.62) & -1.20 (1.08) & 0.39 (0.77) & -0.66 (0.76) \\
 & $5$ & -0.73 (0.35) & -0.32 (0.44) & -0.77 (0.71) & 0.67 (0.56) & -0.80 (0.55) \\
  \cline{2-7}
  median BR$_{\bgamma^{\prime}}$ & 2 & -1.70 (0.53) & -1.41 (0.39) & 2.52 (0.65) & 2.70 (0.66) & 1.61 (0.61) \\
   & 3 &-2.35 (0.63) & 0.34 (0.57) & 1.16 (0.77) & 1.62 (0.77) & -1.12 (1.11) \\
   & 4 &-1.97 (0.55) & 0.60 (0.63) & -1.21 (1.09) & 0.39 (0.77) & -0.66 (0.76) \\
   & 5 & -0.73 (0.35) & -0.32 (0.45) & -0.78 (0.72) & 0.67 (0.56) & -0.80 (0.55) \\
   \bottomrule
\end{tabular}
\end{center}
\end{table}

Table~\ref{estimatemultinomialmodel} gives the ML, mean BR and median BR estimates, along with the corresponding estimated standard errors of the coefficients of model~(\ref{multinomialmodel2}). Table~\ref{estimatemultinomialmodel} shows also results for median BR$_{\bgamma^{\prime}}$, which correspond to the median BR estimates of $\bgamma^{\prime}$ transformed in the $\bgamma$ parameterization. As in logistic regression the mean and median BR estimates are shrunken relative to the maximum likelihood ones with a corresponding shrinkage effect on the estimated standard errors.

The median BR and median BR$_{\bgamma^{\prime}}$ estimates are almost the same, indicating that median BR, in this particular setting, is not  affected by its lack of invariance under linear contrasts.
\begin{table}[t]
\caption{Estimates and estimated standard errors (in parenthesis) of the multinomial regression model (\ref{multinomialmodel2}) for the alligator data in Section~\ref{sec:multinomial} after having the frequencies, and rounding them to the closest integer.}
\begin{center}
\begin{tabular}{lrrrrrr}
\toprule
Method & $c$ & \multicolumn{1}{c}{$\gamma_{c1}$} &
                                                   \multicolumn{1}{c}{$\gamma_{c2}$} & \multicolumn{1}{c}{$\gamma_{c3}$} & \multicolumn{1}{c}{$\gamma_{c4}$} & \multicolumn{1}{c}{$\gamma_{c5}$} \\
  \midrule
ML & $2$ & -1.83 (0.76) & -1.55 (0.59) & 2.66 (0.94) & 2.81 (0.95) & 1.64 (0.87) \\
  & $3$ & -3.39 (1.25) & 1.40 (1.19) & 1.13 (1.29) & 1.44 (1.29) & -$\infty$ (+$\infty$)\\
  & $4$ & -2.31 (0.86) & 0.66 (1.03) & -$\infty$ (+$\infty$) & 0.58 (1.16) & -0.78 (1.29)  \\
  & $5$ & -0.82 (0.49) & -0.04 (0.67) & -1.35 (1.18) & 0.28 (0.81) & -1.25 (0.88)  \\
   \cline{2-7}
  mean BR & $2$ & -1.64 (0.72) & -1.43 (0.59) & 2.40 (0.91) & 2.54 (0.92) & 1.46 (0.84) \\
  & $3$ & -2.76 (1.00) & 1.08 (0.96) & 0.93 (1.15) & 1.22 (1.15) & -1.24 (1.71) \\
  & $4$& -2.02 (0.78) & 0.55 (0.90) & -1.30 (1.70) & 0.57 (1.08) & -0.57 (1.12) \\
  & $5$ & -0.76 (0.49) & -0.03 (0.66) & -1.03 (1.06) & 0.29 (0.81) & -1.08 (0.84) \\
   \cline{2-7}
  median BR & $2$ & -1.76 (0.74) & -1.45 (0.59) & 2.48 (0.93) & 2.62 (0.93) & 1.54 (0.86) \\
  & $3$ & -3.00 (1.08) & 1.23 (1.03) & 1.02 (1.18) & 1.31 (1.18) & -2.04 (2.45) \\
  & $4$ & -2.15 (0.81) & 0.59 (0.95) & -2.17 (2.49) & 0.56 (1.11) & -0.67 (1.19) \\
  & $5$ & -0.79 (0.49) & -0.04 (0.66) & -1.19 (1.11) & 0.28 (0.81) & -1.16 (0.86) \\
   \cline{2-7}
    median BR$_{\bgamma^{\prime}}$ &2 & -1.74 (0.74) & -1.45 (0.58) & 2.50 (0.92) & 2.64 (0.93) & 1.54 (0.85) \\
   & 3 & -3.12 (1.14) & 1.24 (1.08) & 1.03 (1.24) & 1.32 (1.24) & -2.05 (2.61) \\
   & 4& -2.15 (0.81) & 0.60 (0.95) & -2.20 (2.51) & 0.55 (1.11) & -0.67 (1.19) \\
  &5 & -0.79 (0.49) & -0.03 (0.66) & -1.20 (1.11) & 0.27 (0.81) & -1.16 (0.86) \\
   \bottomrule
\end{tabular}
\end{center}
\label{estimatemultinomialmodel2}
\end{table}
The differences between the three methods are more notable when the observed counts are divided by two, as can be seen in Table~\ref{estimatemultinomialmodel2}. In this case, data separation results in two of the ML estimates being infinite. This can generally happen with positive probability when data are sparse or when there are large covariate effects \citep{albert+anderson:1984}. As is the case for logistic regression (see Section~\ref{sec:birthweight}), both mean and median BR deliver finite estimates for all parameters. The finiteness of the mean BR estimates has also been observed in \citet{bull+mak+greenwood:2002}.

In order to better assess the properties of the estimators considered in Table~\ref{estimatemultinomialmodel} and Table~\ref{estimatemultinomialmodel2}, we designed a simulation study where the multinomial totals for each covariate setting in the alligator food choice data set are progressively increased as a fraction of their observed values. Specifically, we consider the sets of multinomial totals $\{r m_1, \ldots, r m_n\}$ for $r \in \{0.5, 0.75, 1, 1.25, 1.5, 1.75, 2, 3, 4, 5\}$, where $m_i$ $(i = 1, \ldots, n)$ is the observed multinomial total for the $i$th combination of covariate values. For each value of $r$, we simulate 10\,000 data sets from the ML fit of model (\ref{multinomialmodel2}) given in Table \ref{estimatemultinomialmodel} and then compare the mean BR, median BR and median BR$_{\bgamma^{\prime}}$ estimators in terms of relative bias and percentage of underestimation. The ML estimator is not considered in the comparison because the probability of infinite estimates is very high, ranging from $1.3$\% for $r = 5$ up to $76.4$\% for $r = 0.5$. In contrast, mean BR and median BR produced finite estimates for all data sets and $r$ values considered.
\begin{figure}[h]
\centering
\includegraphics[scale=0.85]{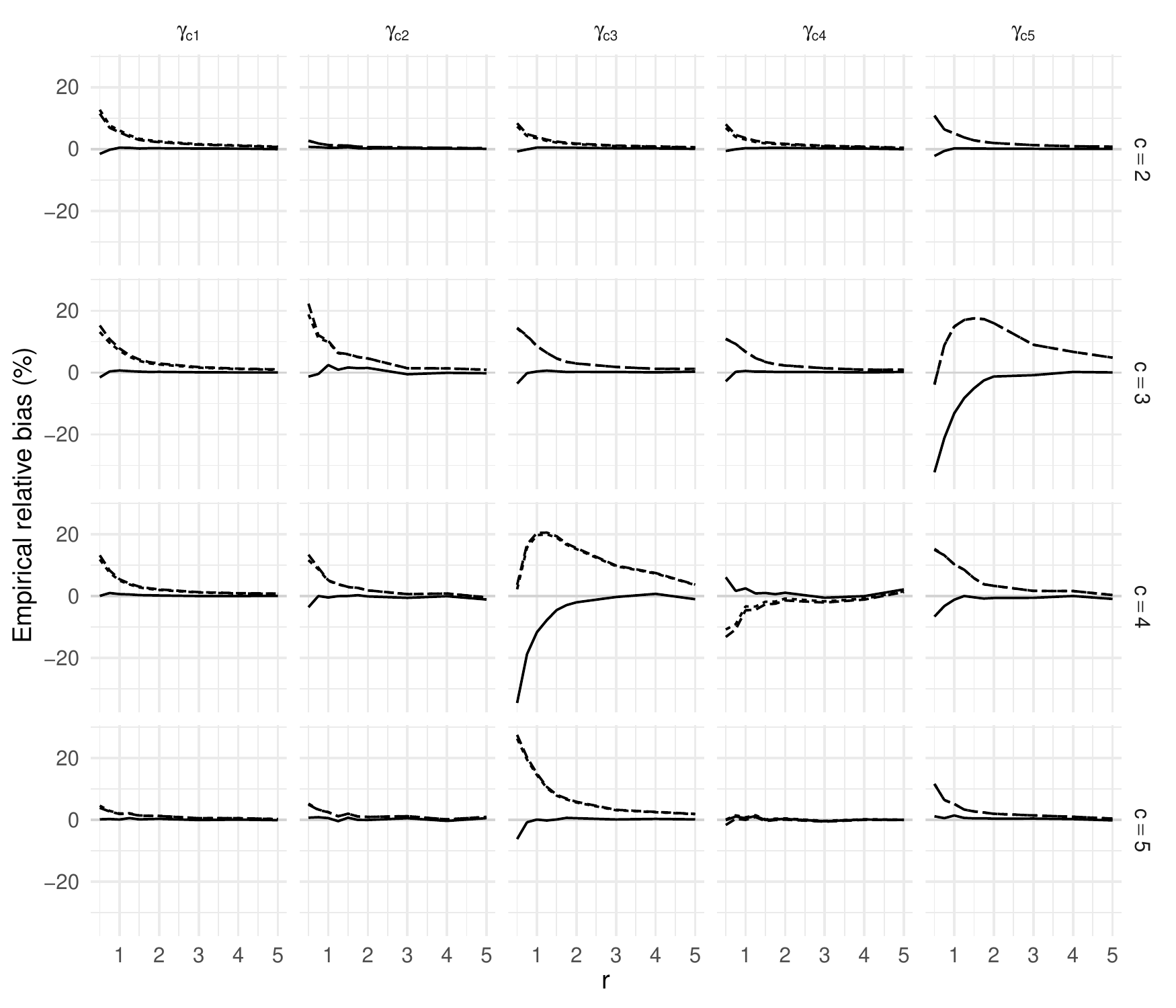}
\caption{Empirical relative bias based on 10\,000 simulated samples from the ML fit of model (\ref{multinomialmodel2}) given in Table \ref{estimatemultinomialmodel}, for each for
  $r \in \{0.5, 0.75, 1, 1.25, 1.5, 1.75, 2, 3, 4, 5\}$. The curves correspond to the mean BR (solid), median BR (dashed), and median BR$_{\bgamma^{\prime}}$ (long-dashed) estimators. The grey horizontal line is at zero.} \label{multinomialbias}
\end{figure}
\begin{figure}[h]
\centering
\includegraphics[scale=0.85]{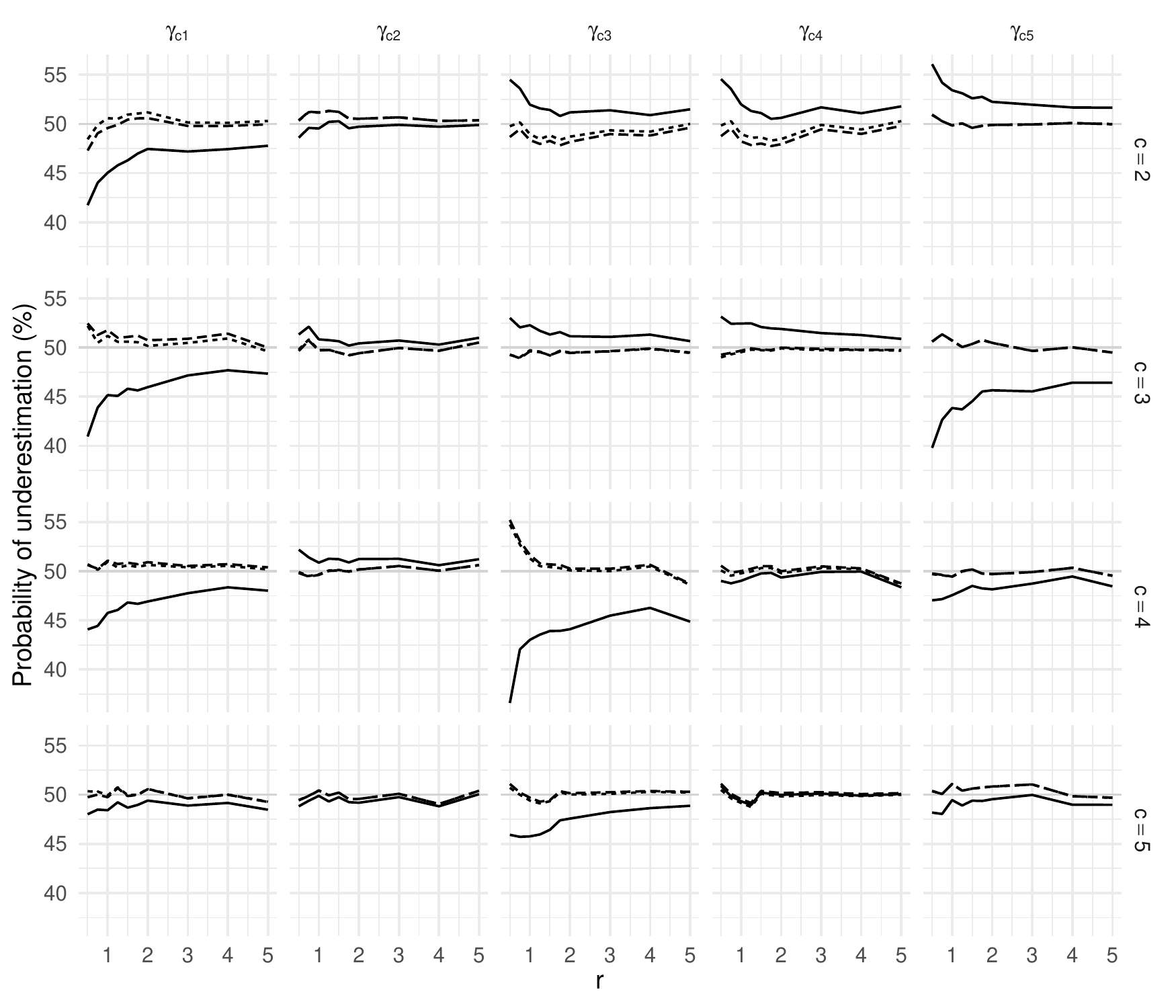}
\caption{Empirical probability of underestimation based on 10\,000 simulated samples from the ML fit of model (\ref{multinomialmodel2}) given in Table \ref{estimatemultinomialmodel}, for each $r \in \{0.5, 0.75, 1, 1.25, 1.5, 1.75, 2, 3, 4, 5\}$. The curves correspond to the mean BR (solid), median BR (dashed), and median BR$_{\bgamma^{\prime}}$ (long-dashed) estimators. The grey horizontal line is at 50.}\label{multinomialpu}
\end{figure}
Figures \ref{multinomialbias} and \ref{multinomialpu} show the relative bias and the percentage of underestimation, respectively, for each parameter as a function of $r$. Overall, mean BR is preferable in terms of mean bias, while median BR achieves better median centering for all the parameters. We note that even median BR$_{\bgamma^{\prime}}$ has bias and probabilities of underestimation very close to those obtained directly under the $\bgamma$ parameterization. This confirms the indications from the observed data that, even if not granted by the theory, median BR is close to invariant under contrasts in the current model setting. As expected, the frequentist properties of the three estimators converge to what we expect from standard ML asymptotics as $r$ increases. In particular, the bias converges to 0 and the percentage of underestimation to $50$\%.


\section{Discussion}
\label{sec:discussion}
Fisher orthogonality \citep{cox+reid:1987} of the mean and dispersion parameters dictates that the mixed approach to bias reduction is valid also for generalized linear models with dispersion covariates in \citet{smyth:1989}, and that estimation can be done by direct generalisation of the IWLS iterations in (\ref{adjusted_iwls}) and (\ref{adjusted_iwls_median_br}), for mean and median bias reduction, respectively.

Inference and model comparison has been based on Wald-type statistics. For special models, it is possible to form penalized likelihood ratio statistics based on the penalized log-likelihood that corresponds to the adjusted scores. A prominent example is logistic regression where the mean bias-reducing adjusted score is the gradient of the log-likelihood penalized by the logarithm of the Jeffreys' prior \citep[see,][where the profiles of the penalized log-likelihood are used for inference]{heinze+schemper:2002}. In that case, the estimator from mean BR  coincides with the mode of the posterior distribution obtained using the Jeffreys' prior \citep[see also][]{ibrahim+laud:1991}. The same happens for Poisson log-linear models and for multinomial baseline category models. Even when a penalized log-likelihood corresponding to adjusted scores is not available \citep[see, Theorem~1 in][ for necessary and sufficient conditions for the existence of mean bias-reducing penalized likelihoods for generalized linear models]{kosmidis+firth:2009}, the adjustments to the score can however be seen as model-based penalties to the inferential quantities for maximum likelihood. In this sense, the adjustments introduce some implicit regularization to the estimation problem, which is just enough to achieve mean or median BR.

In this framework, a general alternative to Wald-type statistics is score-type statistics with known asymptotic distributions, which can be readily defined as in \citet{lindsay+qu:2003}. Let $(\bbeta^\top, \phi)^\top = (\bpsi^\top, \blambda^\top)^\top$, with $\dim(\bpsi) = p_1$ and $\dim(\blambda)= p - p_1$, $\bi^{\bpsi\bpsi}(\bpsi, \blambda)$ be a $p_1 \times p_1$ matrix collecting the rows and columns of $\{\bi(\bpsi, \blambda)\}^{-1}$ corresponding to $\bpsi$, and $\blambda_{\bpsi}^*$ the estimator of $\blambda$ resulting from the solution of the mean bias-reducing adjusted score equations on $\blambda$ for fixed $\bpsi$. Since the scores have an asymptotic normal distribution with mean zero and variance-covariance matrix $\bi(\bpsi, \blambda)$ and the mean bias-reducing adjustment is of order $O(1)$,
\begin{equation}
  \label{adjusted_score_statistics}
  \{\bs_{\bpsi}(\bpsi, \blambda_{\bpsi}^*) + \bA^*_{\bpsi}(\bpsi,
  \blambda_{\bpsi}^*)\}^\top \bi^{\bpsi\bpsi}(\bpsi, \blambda_{\bpsi}^*) \left\{\bs_{\bpsi}(\bpsi, \blambda_{\bpsi}^*) + \bA^*_{\bpsi}(\bpsi, \blambda_{\bpsi}^*) \right\}
\end{equation}
has an asymptotic null $\chi^2_{p_1}$ distribution. The same result holds for median BR, by replacing $\blambda_{\bpsi}^*$ and $\bA_{\bpsi}^*$ with $\blambda_{\bpsi}^\dagger$ and $\bA_{\bpsi}^\dagger$. The adjusted score statistic can then be used for constructing confidence intervals and regions and testing hypotheses on any set of parameters of the generalized linear models, including constructing tables similar to analysis of deviance tables for maximum likelihood.

Finally, as is illustrated in the example of Section \ref{sec:infertility} and shown in \citet{lunardon:2018} and \citet{kenne+salvan+sartori:2017}, mean BR and median BR can be particularly effective for inference about a low-dimensional parameter of interest in the presence of high-dimensional nuisance parameters, while providing, at the same time, improved estimates of the nuisance parameters.


\section{Supplementary material}

The supplementary material includes R code and a report to fully
reproduce all numerical results and figures in the paper.

\section{Acknowledgements}
Ioannis Kosmidis was supported by The Alan Turing Institute under the
EPSRC grant EP/N510129/1 (Turing award number TU/B/000082). Euloge
Clovis Kenne Pagui and Nicola Sartori were supported by the Italian
Ministry of Education under the PRIN 2015 grant 2015EASZFS\_003, and
by the University of Padova (PRAT 2015 CPDA153257).

\section*{Appendix}
\label{appendix}

\subsection*{Proof of Theorem~\ref{the:closer}}

\begin{proof}
  Since $\hat\phi < \phi^* < \phi^\dagger$ and
  $z_{1-\alpha/2} < t_{n-p;1-\alpha/2}$ we have
  $\hat{I}_{1-\alpha} \subset I_{1-\alpha}^* \subset I^E_{1-\alpha}$
  and $ I_{1-\alpha}^* \subset {I}_{1-\alpha}^\dagger$ for any $n-p \ge 1$ and $\alpha \in (0,1)$. We also have
  $I_{1-\alpha}^\dagger \subset {I}_{1-\alpha}^E$ if $g(\nu,\alpha)=
  \{(\nu-2/3)/\nu\}^{1/2}\,
  t_{\nu;1-\alpha/2}-z_{1-\alpha/2} >0$. For fixed natural $\nu\ge 1$, the function $g(\nu,\alpha)$  is positive when $\alpha \rightarrow 0^+$ and has only one zero in $\tilde\alpha(\nu)$. Hence, the condition is satisfied for $\alpha < \tilde\alpha(\nu)$. Moreover, it can be seen numerically that $\tilde\alpha(\nu)$ increases with $\nu$, having a minimum in $\tilde\alpha(1)=0.35562$.
  
Even when $I_{1-\alpha}^E \subset {I}_{1-\alpha}^\dagger$, when $\nu > 1$, the
  absolute difference between the length of the intervals
  $I_{1-\alpha}^\dagger$ and ${I}_{1-\alpha}^E$ is smaller than the
  corresponding difference for $I_{1-\alpha}^*$ and
  ${I}_{1-\alpha}^E$, for any $\alpha>0$. Indeed, this is true provided that the function $h(\nu,\alpha)=2 t_{\nu;1-\alpha/2}/\sqrt{\nu} - z_{1-\alpha/2}/\sqrt{\nu-2/3}- z_{1-\alpha/2}/\sqrt{\nu}$ is positive. This is verified because, for fixed $\nu>1$, $h(\nu,\alpha)$ is a monotonic decreasing function in $\alpha$, converging to $0^+$ as $\alpha \rightarrow 1^-$. On the other hand, if $\nu=1$, $h(\nu,\alpha)$ is positive for $\alpha < 0.62647$ and negative otherwise. 
  
\end{proof}



\section*{Supplementary material}

# Supplementary Material for
# Mean and median bias reduction in generalized linear models

Ioannis Kosmidis, Euloge Clovis Kenne Pagui and Nicola Sartori

January 11, 2019

## 1 Introduction

The current report reproduces the numerical results and figures in the main text. The outputs have been produced using R version 3.5.1 (R Core Team, 2018) and the package `brglm2` (Kosmidis, 2018).

The code chunk below checks for any installed versions and installs the `brglm2` R package for mean and median bias reduction in generalized linear models, and loads the R packages that are used for the reproduction of numerical results in the main text.

```r
brglm2_version <- try(packageVersion("brglm2"), silent = TRUE)
if (!inherits(brglm2_version, "try-error")) {
    if (brglm2_version < "0.2") {
        devtools::install_github("ikosmidis/brglm2")
    }
}
library("brglm2")
library("MASS")
library("survival")
library("ggplot2")
library("dplyr")
```

We also provide code to reproduce all model fits and simulation results in the main text. The R scripts to carry out the simulation experiments, and the results from those are provided in the `glmbias_code+results.zip` archive. `res_dir` is the directory where the contents of the archive are and needs to be set appropriately.

```r
res_dir <- "~/Repositories/glmbias/text/supplementary/glmbias_code+results"
```

## 2 Gamma regression model for blood clotting times

This section provides the R code that reproduces the numerical results of Example 1 and Section 5.2 of the paper.

The code chunk below reproduces the reported estimates in Example 1.1 and Table 5 of the main text

```r
## clotting data
clotting <- data.frame(
  u = c(5,10,15,20,30,40,60,80,100, 5,10,15,20,30,40,60,80,100),
  conc = c(118,58,42,35,27,25,21,19,18,69,35,26,21,18,16,13,12,12),
  lot = factor(c(rep(1, 9), rep(2, 9))))
## maximum likelihood fit
clot_ML <- glm(conc~lot*log(u), data = clotting, family = Gamma(link="log"),
method="brglmFit",type="ML")
d2afuns <- enrichwith::enrich(clot_ML$family, with = c("d1afun", "d2afun",
```

```r
                           "d3afun", "d1variance"))$d2afun
weights <- as.vector(clot_ML$prior.weights)
clot_ml <- summary(clot_ML)
zetas <- -weights/clot_ml$dispersion
resml <- cbind(t(clot_ml$coefficients[,1:2]),
               precision=c(clot_ml$dispersion,
                           sqrt((2*clot_ml$dispersion^4)/
                                sum(weights^2*d2afuns(zetas), na.rm = TRUE))) )
round(resml, 3)

##            (Intercept)   lot2 log(u) lot2:log(u) precision
## Estimate         5.503 -0.584 -0.602       0.034     0.017
## Std. Error       0.161  0.228  0.047       0.066     0.006

## mean bias-reduced fit
clot_meanBR <- update(clot_ML, method = "brglmFit", type = "AS_mean")
clot_meanBR <- summary(clot_meanBR)
zetas <- -weights/clot_meanBR$dispersion
resmeanBR <- cbind(t(clot_meanBR$coefficients[,1:2]),
                   precision=c(clot_meanBR$dispersion,
                               sqrt((2*clot_meanBR$dispersion^4)/
                                    sum(weights^2*d2afuns(zetas), na.rm = TRUE))))
round(resmeanBR, 3)

##            (Intercept)   lot2 log(u) lot2:log(u) precision
## Estimate         5.507 -0.584 -0.602       0.034     0.022
## Std. Error       0.183  0.258  0.053       0.075     0.007

## median bias-reduced fit
clot_medianBR <- update(clot_ML, method = "brglmFit", type = "AS_median")
clot_medianBR <- summary(clot_medianBR)
zetas <- -weights/clot_medianBR$dispersion
resmedianBR <- cbind(t(clot_medianBR$coefficients[,1:2]),
                     precision=c(clot_medianBR$dispersion,
                                 sqrt((2*clot_medianBR$dispersion^4)/
                                      sum(weights^2*d2afuns(zetas), na.rm = TRUE))))
round(resmedianBR, 3)

##            (Intercept)   lot2 log(u) lot2:log(u) precision
## Estimate         5.505 -0.584 -0.602       0.034     0.024
## Std. Error       0.187  0.265  0.054       0.077     0.008

## mixed bias-reduced fit
clot_mixed <- update(clot_ML, method = "brglmFit", type = "AS_mixed")
clot_mixed <- summary(clot_mixed)
zetas <- -weights/clot_mixed$dispersion
resmixed <- cbind(t(clot_mixed$coefficients[,1:2]),
                  precision=c(clot_mixed$dispersion,
                              sqrt((2*clot_mixed$dispersion^4)/
                                   sum(weights^2*d2afuns(zetas), na.rm = TRUE))))
round(resmixed, 3)

##            (Intercept)   lot2 log(u) lot2:log(u) precision
## Estimate         5.507 -0.584 -0.602       0.034     0.024
## Std. Error       0.187  0.265  0.054       0.077     0.008
```

The following code chunk uses the image file `clotting_simulation_results.rda` to reproduce Table 1 and Table 6 of the main text. `clotting_simulation_results.rda` results by running the script `glmbias_gamma_simulation.R` which is available in the supplementary code archive.

```r
load(paste(res_dir, "clotting_simulation_results.rda", sep = "/"))
out <- rbind(mle, meanBR, medianBR, meanmixed)
rmse <- sqrt(out[,1]^2+out[,2]^2)
out[,2] <- rmse
round(out * 100, 2)

##                [,1]  [,2]  [,3]  [,4]  [,5]  [,6]
## (Intercept)   -0.33 16.15  0.04 50.42 12.87 89.26
## lot2           0.36 23.09  0.02 49.61 18.46 88.87
## log(u)         0.06  4.69  0.01 49.73  3.74 89.62
## lot2:log(u)  -0.11  6.71  0.03 50.51  5.36 88.78
## dispersion   -0.38  0.65 54.13 78.77  0.55 69.26
## (Intercept)  -0.04 16.15  0.00 49.65 12.87 93.12
## lot2          0.36 23.09  0.02 49.59 18.46 92.69
## log(u)        0.02  4.69  0.00 49.92  3.74 93.08
## lot2:log(u) -0.11  6.71  0.03 50.50  5.36 92.26
## dispersion   0.00  0.67  0.00 55.00  0.53 86.22
## (Intercept) -0.15 16.15  0.01 49.93 12.87 93.67
## lot2         0.36 23.09  0.02 49.60 18.46 93.27
## log(u)       0.03  4.69  0.01 49.88  3.74 93.73
## lot2:log(u) -0.11  6.71  0.03 50.50  5.36 93.05
## dispersion   0.09  0.71  1.67 49.99  0.55 88.44
## (Intercept) -0.02 16.15  0.00 49.65 12.87 93.66
## lot2         0.36 23.09  0.02 49.59 18.46 93.28
## log(u)       0.02  4.69  0.00 49.95  3.74 93.71
## lot2:log(u) -0.11  6.71  0.03 50.50  5.36 93.06
## dispersion   0.09  0.71  1.68 49.93  0.55 88.45
```

## 3 Mixed adjustments for dispersion models

This section provides the R code that reproduces the figures in Table 4 of the main text.

```r
load(paste(res_dir, "mixed_invariance_simulation_results.rda", sep = "/"))
round(cbind(eps_cont, cont_probs, eps_phi, disp_probs), digits = 3)

##      eps_cont ML mean BR median BR mixed eps_phi ML mean BR median BR
## [1,]     0.01  0         0    0.656     0    0.02  0     0.978         0
## [2,]     0.02  0         0    0.162     0    0.04  0     0.771         0
## [3,]     0.03  0         0    0.034     0    0.06  0     0.454         0
## [4,]     0.04  0         0    0.010     0    0.08  0     0.181         0
## [5,]     0.05  0         0    0.003     0    0.10  0     0.061         0
##      mixed
## [1,]     0
## [2,]     0
## [3,]     0
## [4,]     0
## [5,]     0
```

## 4 Logistic regression for infant birth weights

This section provides the R code that reproduces the numerical results of Section 5.3 of the paper.

The code chunk below reproduces the figures in Table 7 of the main text.

```r
## Prepare the birth weight data set
bwt <- with(birthwt, {
    age <- age
    racewhite <- ifelse(race==1,1,0)
```

```r
    smoke <- smoke
    ptl <- ifelse(ptl>0,1,0)
    ptd <- factor(ptl > 0)
    ht <- ht
    loglwt <- log(lwt)
    data.frame(normwt = 1-low, age, racewhite, smoke, ptl,ht,loglwt,ftv)
})
bwt <- subset(bwt, subset = (ftv==0), select= -c(ftv))
## maximum likelihood fit
bwt_ml <- glm(normwt ~ ., family = binomial, data = bwt)
round(t(summary(bwt_ml)$coefficients[, 1:2]), 3)

##             (Intercept)    age racewhite  smoke    ptl     ht loglwt
## Estimate         -8.496 -0.067     0.690 -0.560 -1.603 -1.211  2.262
## Std. Error        5.826  0.053     0.566  0.576  0.697  0.924  1.252

## mean bias-reduced fit ##
bwt_br <- update(bwt_ml, method = "brglmFit", type = "AS_mean")
round(t(summary(bwt_br)$coefficients[, 1:2]), 3)

##             (Intercept)    age racewhite  smoke    ptl     ht loglwt
## Estimate         -7.401 -0.061     0.622 -0.531 -1.446 -1.104  1.998
## Std. Error        5.664  0.052     0.552  0.564  0.680  0.901  1.216

## median bias-reduced fit ##
bwt_mbr <- update(bwt_ml, method = "brglmFit", type = "AS_median")
round(t(summary(bwt_mbr)$coefficients[, 1:2]), 3)

##             (Intercept)    age racewhite  smoke    ptl     ht loglwt
## Estimate         -7.641 -0.062     0.638 -0.538 -1.481 -1.134  2.059
## Std. Error        5.717  0.053     0.557  0.568  0.681  0.906  1.228
```

The code chunk below uses the image file `birth_weight_simulation_results.rda` to reproduce the figures in Table 8 of the main text. `birth_weight_simulation_results.rda` results by running the script `glmbias_logistic_simulation.R` which is available in the supplementary code archive.

```r
load(paste(res_dir, "birth_weight_simulation_results.rda", sep = "/"))
t(round(bias.beta, 2))

##       [,1]  [,2] [,3]  [,4]  [,5]  [,6] [,7]
## ml   -1.42 -0.01 0.09 -0.03 -0.20 -0.12 0.34
## br   -0.08  0.00 0.01  0.00 -0.01  0.00 0.02
## mbr  -0.38  0.00 0.03 -0.01 -0.07 -0.04 0.09

t(round(bias.psi, 2))

##        [,1] [,2] [,3] [,4] [,5] [,6]  [,7]
## ml   183.50    0 0.75 0.12 0.02 0.18 57.50
## br    47.17    0 0.41 0.11 0.05 0.17 18.75
## mbr   56.66    0 0.50 0.11 0.04 0.21 23.74

t(round(sqrt(mse.beta),2))

##      [,1] [,2] [,3] [,4] [,5] [,6] [,7]
## ml   6.86 0.06 0.66 0.66 0.82 1.11 1.49
## br   5.94 0.05 0.58 0.59 0.72 0.94 1.28
## mbr  6.11 0.06 0.60 0.61 0.78 1.01 1.32

t(round(PU * 100, 1))

##      [,1] [,2] [,3] [,4] [,5] [,6] [,7]
## ml   56.1 53.2 46.4 51.4 57.4 53.5 43.1
## br   48.2 49.2 51.3 49.6 48.1 48.9 52.2
## mbr  50.0 49.6 49.9 49.9 50.6 50.3 50.0
```

4

```
t(round(coverage * 100, 1))

##     [,1] [,2] [,3] [,4] [,5] [,6] [,7]
## ml  94.8 94.8 94.5 94.7 96.4 96.6 94.6
## br  96.3 96.2 96.0 96.2 97.2 98.1 96.1
## mbr 96.1 96.0 95.8 95.9 97.0 97.8 96.0
```

# 5 Logistic regression for the link between sterility and abortion

This section provides the R code that reproduces the numerical results of Section 5.4 of the paper.

The code chunk below reproduces the results in Table 9 of the main text.

```
data("infert")
infert0 <- infert
infert0$spontaneous <- factor(infert0$spontaneous)
infert0$induced <- factor(infert0$induced)
infert0$stratum <- factor(infert0$stratum)

## glm with subject-specific parameter
mml <- glm(case ~ -1 + stratum + spontaneous + induced,
           family = binomial, data = infert0)
sml <- summary(mml)$coef[83 + (1:4), ]
## conditional likelihood
mcl <- clogit(case ~ spontaneous + induced + strata(stratum),
              data = infert0, method = "exact")
scl <- summary(mcl)$coef[, c(1, 3)]
## mean bias reduction
mbr <- update(mml, method = 'brglmFit', type = 'AS_mean')
sbr <- summary(mbr)$coef[83 + (1:4), ]
## median bias reduction
mmbr <- update(mml, method = 'brglmFit', type = 'AS_median')
smbr <- summary(mmbr)$coef[83 + (1:4), ]

round(rbind(t(sml[1:4, 1:2]),
            t(scl[1:4, 1:2]),
            t(sbr[1:4, 1:2]),
            t(smbr[1:4, 1:2])), 3)

##              spontaneous1 spontaneous2 induced1 induced2
## Estimate            3.268        6.441    2.112    4.418
## Std. Error          0.592        0.955    0.587    0.948
## coef                2.044        3.935    1.386    2.819
## se(coef)            0.453        0.725    0.463    0.735
## Estimate            2.055        3.954    1.305    2.714
## Std. Error          0.472        0.708    0.474    0.744
## Estimate            2.083        3.997    1.330    2.760
## Std. Error          0.478        0.713    0.482    0.754
```

# 6 Primary food choices of alligators

This section provides the R code that reproduces the numerical results of Section 6.3 of the paper.

The code chunk below reproduces the results in Table 10 of the main text.

```r
data("alligators", package = "brglm2")
all_ml <- brmultinom(foodchoice ~ size + lake , weights = freq,
                     data = alligators,
                     ref = 1,
                     type = "ML")
all_mean <- update(all_ml, type = "AS_mean")
all_median <- update(all_ml, type = "AS_median")
agresti_contrasts <- list(
    lake = contr.treatment(levels(alligators$lake), base = 4),
    size = contr.treatment(levels(alligators$size), base = 2))
## Contrasts matrix
mat <- cbind(c(1, 1, 1, 0, 0), c(0, -1, 0, 0, 0), c(0, 0, -1, 1, 0),
             c(0, 0, -1, 0, 1), c(0, 0, -1, 0, 0))
all_median_agresti <- brmultinom(foodchoice ~ size + lake , weights = freq,
                                 data = alligators,
                                 contrasts = agresti_contrasts,
                                 ref = 1,
                                 type = "AS_median")
median_gamma <- summary(all_median_agresti)$coefficients %*% mat
colnames(median_gamma) <- colnames(coef(all_ml))
kmat <- kronecker(diag(4), mat)
median_gamma_se <- matrix(sqrt(diag(t(kmat) %*% vcov(all_median_agresti) %*% kmat)),
                          ncol = 5, byrow = TRUE)
dimnames(median_gamma_se) <- dimnames(coef(all_ml))
round(summary(all_ml)$coefficients, 2)

##              (Intercept) size>2.3 lakeOklawaha lakeTrafford lakeGeorge
## Invertebrate       -1.75    -1.46         2.60         2.78       1.66
## Reptile            -2.42     0.35         1.22         1.69      -1.24
## Bird               -2.03     0.63        -1.35         0.39      -0.70
## Other              -0.75    -0.33        -0.82         0.69      -0.83

round(summary(all_ml)$standard.errors, 2)

##              (Intercept) size>2.3 lakeOklawaha lakeTrafford lakeGeorge
## Invertebrate        0.54     0.40         0.66         0.67       0.61
## Reptile             0.64     0.58         0.79         0.78       1.19
## Bird                0.56     0.64         1.16         0.78       0.78
## Other               0.35     0.45         0.73         0.56       0.56

round(summary(all_mean)$coefficients, 2)

##              (Intercept) size>2.3 lakeOklawaha lakeTrafford lakeGeorge
## Invertebrate       -1.65    -1.40         2.46         2.64       1.56
## Reptile            -2.25     0.32         1.12         1.58      -0.98
## Bird               -1.90     0.58        -1.04         0.40      -0.62
## Other              -0.72    -0.31        -0.72         0.67      -0.78

round(summary(all_mean)$standard.errors, 2)

##              (Intercept) size>2.3 lakeOklawaha lakeTrafford lakeGeorge
## Invertebrate        0.52     0.40         0.65         0.66       0.60
## Reptile             0.61     0.56         0.76         0.75       1.02
## Bird                0.54     0.61         1.01         0.76       0.74
## Other               0.35     0.44         0.71         0.56       0.55

round(summary(all_median)$coefficients, 2)

##              (Intercept) size>2.3 lakeOklawaha lakeTrafford lakeGeorge
## Invertebrate       -1.71    -1.41         2.51         2.69       1.61
## Reptile            -2.33     0.34         1.16         1.62      -1.12
## Bird               -1.96     0.60        -1.20         0.39      -0.66
## Other              -0.73    -0.32        -0.77         0.67      -0.80
```

```r
round(summary(all_median)$standard.errors, 2)

##              (Intercept) size>2.3 lakeOklawaha lakeTrafford lakeGeorge
## Invertebrate        0.53     0.40         0.65         0.67       0.61
## Reptile             0.62     0.57         0.77         0.76       1.10
## Bird                0.54     0.62         1.08         0.77       0.76
## Other               0.35     0.44         0.71         0.56       0.55

round(median_gamma, 2)

##              (Intercept) size>2.3 lakeOklawaha lakeTrafford lakeGeorge
## Invertebrate       -1.70    -1.41         2.52         2.70       1.61
## Reptile            -2.35     0.34         1.16         1.62      -1.12
## Bird               -1.97     0.60        -1.21         0.39      -0.66
## Other              -0.73    -0.32        -0.78         0.67      -0.80

round(median_gamma_se, 2)

##              (Intercept) size>2.3 lakeOklawaha lakeTrafford lakeGeorge
## Invertebrate        0.53     0.39         0.65         0.66       0.61
## Reptile             0.63     0.57         0.77         0.77       1.11
## Bird                0.55     0.63         1.09         0.77       0.76
## Other               0.35     0.45         0.72         0.56       0.55
```

The figures in Table 11 of the main text are obtained as

```r
all_half <- within(alligators, freq <- round(freq/2))
all_ml <- update(all_ml, data = all_half)
all_mean <- update(all_ml, type = "AS_mean")
all_median <- update(all_ml, type = "AS_median")
all_median_agresti <- update(all_median_agresti, data = all_half)
median_gamma <- summary(all_median_agresti)$coefficients %*% mat
colnames(median_gamma) <- colnames(coef(all_ml))
median_gamma_se <- matrix(sqrt(diag(t(kmat) %*% vcov(all_median_agresti) %*% kmat)),
                          ncol = 5, byrow = TRUE)
dimnames(median_gamma_se) <- dimnames(coef(all_ml))
round(summary(all_ml)$coefficients, 2)

##              (Intercept) size>2.3 lakeOklawaha lakeTrafford lakeGeorge
## Invertebrate       -1.83    -1.55         2.66         2.81       1.64
## Reptile            -3.39     1.40         1.13         1.44    -133.35
## Bird               -2.31     0.66      -133.70         0.58      -0.78
## Other              -0.82    -0.04        -1.35         0.28      -1.25

round(summary(all_ml)$standard.errors, 2)

##              (Intercept) size>2.3 lakeOklawaha lakeTrafford   lakeGeorge
## Invertebrate        0.76     0.59         0.94         0.95         0.87
## Reptile             1.25     1.19         1.29         1.29  22369621.33
## Bird                0.86     1.03  23726566.41         1.16         1.29
## Other               0.49     0.67         1.18         0.81         0.88

round(summary(all_mean)$coefficients, 2)

##              (Intercept) size>2.3 lakeOklawaha lakeTrafford lakeGeorge
## Invertebrate       -1.64    -1.43         2.40         2.54       1.46
## Reptile            -2.76     1.08         0.93         1.22      -1.24
## Bird               -2.02     0.55        -1.30         0.57      -0.57
## Other              -0.76    -0.03        -1.03         0.29      -1.08

round(summary(all_mean)$standard.errors, 2)
```

```
##              (Intercept) size>2.3 lakeOklawaha lakeTrafford lakeGeorge
## Invertebrate        0.72     0.59         0.91         0.92       0.84
## Reptile             1.00     0.96         1.15         1.15       1.71
## Bird                0.78     0.90         1.70         1.08       1.12
## Other               0.49     0.66         1.06         0.81       0.84
```

```r
round(summary(all_median)$coefficients, 2)
```

```
##              (Intercept) size>2.3 lakeOklawaha lakeTrafford lakeGeorge
## Invertebrate       -1.76    -1.45         2.48         2.62       1.54
## Reptile            -3.00     1.23         1.02         1.31      -2.04
## Bird               -2.15     0.59        -2.17         0.56      -0.67
## Other              -0.79    -0.04        -1.19         0.28      -1.16
```

```r
round(summary(all_median)$standard.errors, 2)
```

```
##              (Intercept) size>2.3 lakeOklawaha lakeTrafford lakeGeorge
## Invertebrate        0.74     0.59         0.93         0.93       0.86
## Reptile             1.08     1.03         1.18         1.18       2.45
## Bird                0.81     0.95         2.49         1.11       1.19
## Other               0.49     0.66         1.11         0.81       0.86
```

```r
round(median_gamma, 2)
```

```
##              (Intercept) size>2.3 lakeOklawaha lakeTrafford lakeGeorge
## Invertebrate       -1.74    -1.45         2.50         2.64       1.54
## Reptile            -3.12     1.24         1.03         1.32      -2.05
## Bird               -2.15     0.60        -2.20         0.55      -0.67
## Other              -0.79    -0.03        -1.20         0.27      -1.16
```

```r
round(median_gamma_se, 2)
```

```
##              (Intercept) size>2.3 lakeOklawaha lakeTrafford lakeGeorge
## Invertebrate        0.74     0.58         0.92         0.93       0.85
## Reptile             1.14     1.08         1.24         1.24       2.61
## Bird                0.81     0.95         2.51         1.11       1.19
## Other               0.49     0.66         1.11         0.81       0.86
```

Finally, the code chunk below prepares the data for producing Figure 3 and Figure 4

```r
load(paste(res_dir, "alligator_simulation_results.rda", sep = "/"))
## Prepare plotting data frame
data_ggplot <- rbind(br_mean_results, br_median_results,
                     br_median_gamma_results)
data_ggplot$gamma <- recode(data_ggplot$ind, "(Intercept)" = 1,
                    "size>2.3" = 2, "lakeOklawaha" = 3,
                    "lakeTrafford" = 4, "lakeGeorge" = 5)
data_ggplot$c <- recode(data_ggplot$category, "Invertebrate" = 2,
              "Reptile" = 3, "Bird" = 4, "Other"= 5)
names(data_ggplot) <- c("bias", "parameter", "category", "a_val",
                    "method", "pu", "gamma", "c")
data_ggplot$gamma_label <- paste0("gamma[c", data_ggplot$gamma, "]")
data_ggplot$c_label <- paste0("c==", data_ggplot$c)
```

Figure 3 is the result of

```r
## Relative bias
print(ggplot(data_ggplot) +
    geom_hline(aes(yintercept = 0), col = "grey", alpha = 0.5) +
    geom_line(aes(x = a_val, y = bias, lty = method)) +
    labs(x = expression(r), y = "Empirical relative bias (%)") +
    facet_grid(c_label ~ gamma_label, labeller = label_parsed) +
```

8

```
    theme_minimal() +
    theme(legend.position = "none"))
```

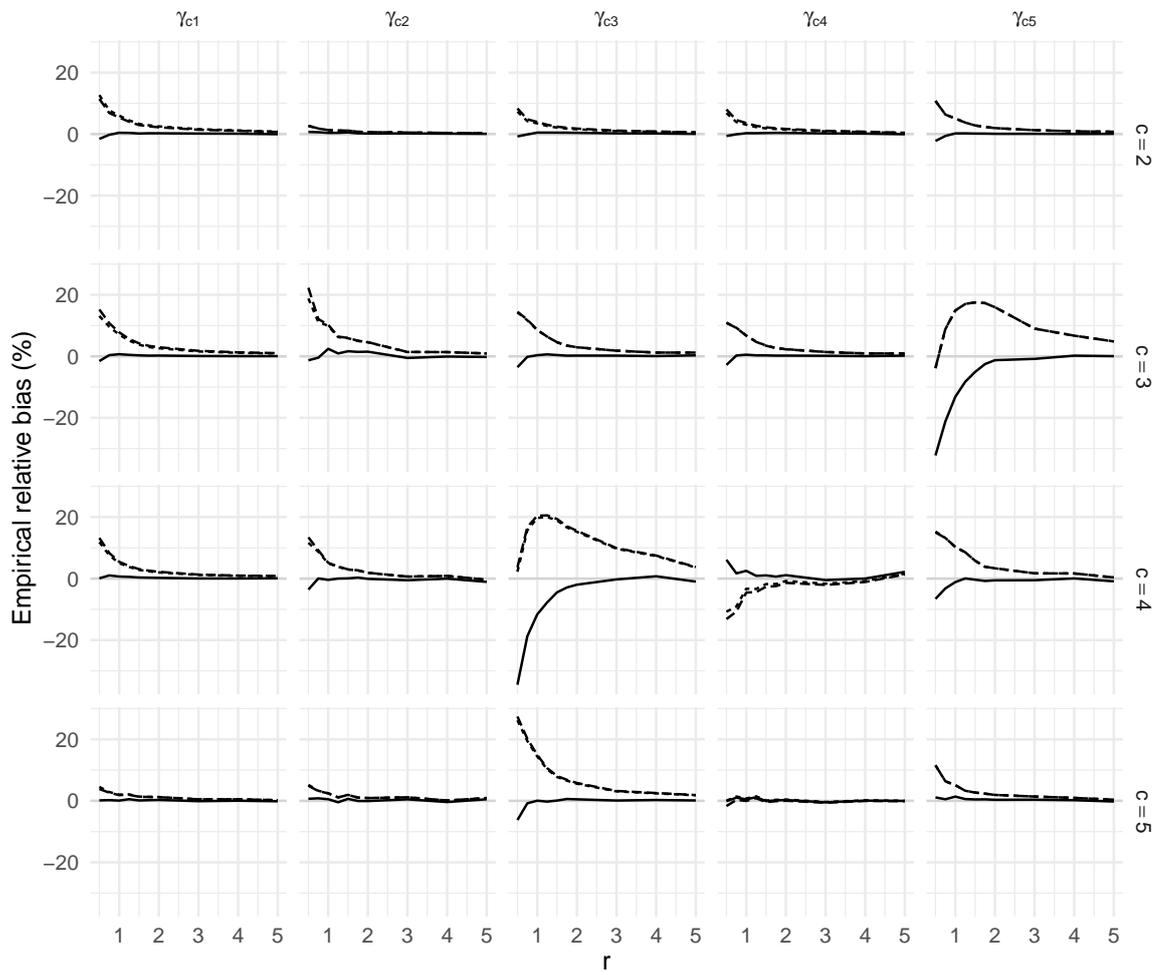

Figure 4 is the result of

```
## Probability of underestimation
print(ggplot(data_ggplot) +
    geom_hline(aes(yintercept = 50), col = "grey", alpha = 0.5) +
    geom_line(aes(x = a_val, y = pu, lty = method)) +
    labs(x = expression(r), y = "Probability of underestimation (%)") +
    facet_grid(c_label ~ gamma_label, labeller = label_parsed) +
    theme_minimal() +
    theme(legend.position = "none"))
```

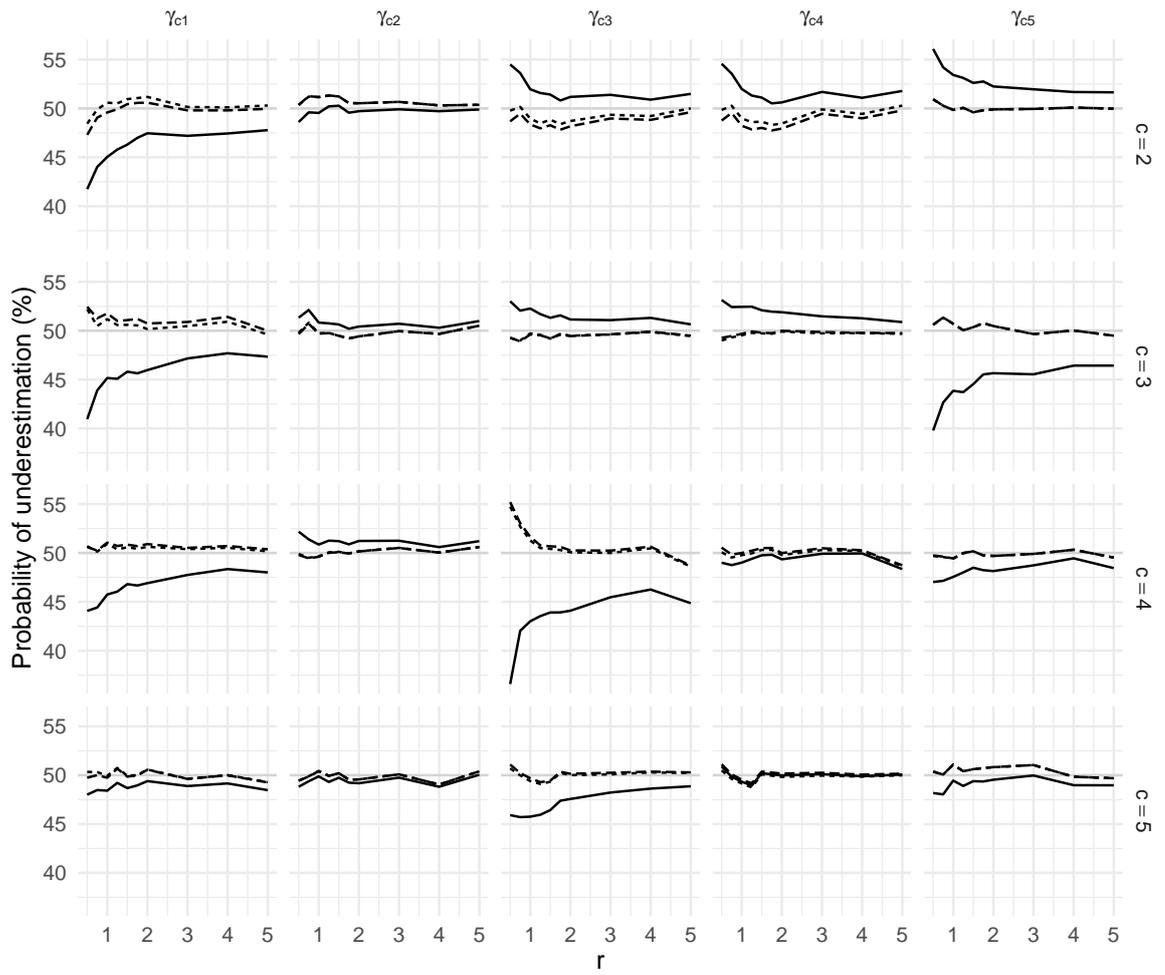



\end{document}